\documentclass[epj]{svjour}
\usepackage{graphicx,epsf,dsfont,amssymb}
\begin{document}
\title{Resilience of public transport networks against attacks}

\author{B. Berche\inst{1}\inst{4}\thanks{berche@lpm.u-nancy.fr}
\and C.\ von
Ferber\inst{2}\inst{3}\thanks{C.vonFerber@coventry.ac.uk}
 \and T. Holovatch\inst{1}\inst{2}\thanks{holtaras@lpm.u-nancy.fr}
 \and Yu. Holovatch\inst{4}\inst{5}\inst{1}\thanks{hol@icmp.lviv.ua}
}                     
%
%
\institute{Statistical Physics Group, P2M Dpt, Institut Jean Lamour,
Nancy Universit\'e, BP 70239, F-54506 Vand\oe uvre les Nancy, France
 \and Applied Mathematics Research Centre, Coventry
University, Coventry CV1 5FB, UK \and
 Physikalisches Institut, Universit\"at Freiburg, D-79104 Freiburg, Germany \and
  Institute for Condensed Matter Physics, National Academy of Sciences of Ukraine, UA--79011 Lviv,
Ukraine \and
 Institut f\"ur Theoretische Physik, Johannes Kepler Universit\"at Linz, A-4040, Linz, Austria
}
\date{Coventry, May 11, 2009}
%
\abstract{The behavior of complex networks under failure or attack
depends strongly on the specific scenario. Of special interest are
scale-free networks, which are usually seen as robust under random
failure but appear to be especially vulnerable to targeted attacks.
In recent studies of public transport networks of fourteen major
cities of the world it was shown that these systems when represented
by appropriate graphs may exhibit scale-free behavior [C. von Ferber
{\em et al.}, Physica A {\bf 380}, 585 (2007), Eur. Phys. J. B  {\bf
68}, 261 (2009)]. Our present analysis, focuses on the effects that
defunct or removed nodes have on the properties of public transport
networks. Simulating different directed attack strategies, we derive
vulnerability criteria that result in minimal strategies with high
impact on these systems.
} 
\maketitle
%
\section{Introduction}\label{I}

The question of resilience or vulnerability of a complex network
\cite{networks} against failure of its parts has, beside purely
academic interest a whole range of important practical implications.
In what follows below any such failure will be called an {\em
attack}. In practice, the origin of the attack and its scenario may
differ to large extent, ranging from random failure, when a node or
a link in a network is removed at random to a targeted destruction,
when the most influential network constituents are removed according
to their operating characteristics. The notion of attack
vulnerability of complex networks originates from studies of
computer networks and was coined to denote the decrease of network
performance as caused by the removal of either nodes or links. The
behavior of a complex network under attack has been observed to
drastically differ from that of regular lattices. Early evidence of
this fact was found in particular for real world networks that show
scale-free behavior: the world wide web and the internet
\cite{Albert00,Tu00}, as well as metabolic \cite{Jeong00}, food web
\cite{Sole01}, and protein \cite{Jeong01} networks. It appeared that
these networks display a high degree of robustness against random
failure. However, if the scenario is changed towards targeted
attacks, the same networks may appear to be especially vulnerable
\cite{Cohen00,Callaway00}.

Essential progress towards a theoretical description of the attack
vulnerability of complex networks is due to the application of the
tools and concepts of percolation phenomena \cite{percolation}. On a
lattice percolation occurs e.g. when at a given concentration of
bonds a spanning cluster appears. This concentration $c_{\rm perc}$
which is determined by an appropriate ensemble average in the
thermodynamic limit is the so-called percolation threshold which is
in general lattice dependent. On a general network the corresponding
phenomenon is the emergence of a giant connected component (GCC)
i.e. a connected subnetwork which in the limit of an infinite
network contains a finite fraction of the network. For a random
graph where given vertices are linked at random this threshold has
been shown to be reached at one bond per vertex \cite{Erdos}.
However the distribution $p(k)$ of the degrees $k$ of vertices in a
random graph is Poissonian. A more general criterion applicable to
networks with given degree distribution $p(k)$ but otherwise random
linking between vertices has been proposed by Molloy and Reed
\cite{Molloy,Cohen00,Callaway00}. For such equilibrium networks a
GCC can be shown to be present if
\begin{equation} \label{1b}
\langle k(k-2) \rangle \geq 0
\end{equation}
with the appropriate ensemble average $\langle \dots \rangle$ over
networks with given degree distribution. Defining the Molloy-Reed
parameter as the ratio of the moments of the degree distribution
\begin{equation} \label{1}
 \kappa^{(k)} \equiv  \langle k^2 \rangle / \langle k \rangle
\end{equation}
the percolation threshold can then be determined by
\begin{equation} \label{1a}
 \kappa^{(k)}=2
 \hspace{1em}\mbox{ at }\hspace{1em} c_{\rm perc}.
\end{equation}
Taken that for scale-free networks the degree distribution obeys
power law scaling
\begin{equation} \label{2}
p(k) \sim k^{-\gamma}
\end{equation}
one finds that the second moment $\langle k^2 \rangle$  diverges for
$\gamma<3$. Thus, the value $\gamma=3$ separates two different
regimes for the percolation on equilibrium scale free networks
\cite{Cohen00}. Indeed, for infinite equilibrium scale-free networks
$\kappa^{(k)}$ (\ref{1}) remains finite for $\gamma>3$, however for
$\gamma<3$ a GCC is found to exist at any concentration of removed
sites: the network appears to be extremely robust to random removal
of nodes. Therefore, observed transitions for real-world systems
\cite{Albert00,Tu00,Jeong00,Sole01,Jeong01} from the theoretical
standpoint may be seen as finite-size effects or resulting from
essential degree-degree correlations. The tolerance of scale-free
networks to intentional attacks (when the highest degree nodes are
removed) was studied in Ref. \cite{Cohen01}. It was shown that even
networks with $\gamma<3$ may be sensitive to intentional attacks.

Obviously, the above theoretical results apply to ideal complex
networks and for ensemble averages and may be confirmed within
certain accuracy when applied to different individual real-world
networks. Not only finite-size effects are the origin of this
discrepancy \cite{Kalisky06}. Furthermore, even networks of similar
type (e.g. of similar node degree distribution and size) may be
characterized by a large variety of other characteristics. While
some of them may have no impact on the percolation properties
\cite{Xulvi-Brunet03}, others do modify their behavior under attack,
as empirically revealed in Ref. \cite{Holme02} for two different
real-world scale-free networks (computer and collaboration
networks). Therefore, an empirical analysis of the behavior of
different real-world networks under attack appears timely and will
allow not only to elaborate scenarios for possible defence
mechanisms of operating networks but also to create strategies of
network constructions, that are robust to attacks of various types.

In this paper, we present results of the analysis of the behavior of
networks of public transport in large cities (public transport
networks, PTNs) and consider attacks by various scenarios. To our
knowledge the resilience of PTNs under attack has so far not been
treated in terms of complex network concepts. Furthermore, in
parallel we analyze a number of complex networks of the same type.
Previous analysis usually focussed on a single instance of a network
of given type \cite{note2}. Our study intends to show that even
within a sample of several networks that were created for the same
purpose, namely PTNs, one may observe essential diversity with
respect to the behavior under attacks of various scenarios.

As we have mentioned above, the attack resilience of a network may
be tested within a variety of different attack scenarios. In a given
one, a list of nodes ordered by decreasing degree may be prepared
for the unperturbed network and the attack successively removes
vertices according to this original list \cite{Barabasi99,Broder00}.
In a slightly different scenario the vertex degrees are recalculated
and the list is reordered after each removal step \cite{Albert00}.
In initial studies only little difference between these two
scenarios was observed \cite{Callaway00}, however further analysis
showed \cite{Holme02,Girvan02} that attacks according to
recalculated lists often turn out to be more harmful than the attack
strategies based on the initial list, suggesting that the network
structure changes as important vertices or edges are removed.  Other
scenarios consider attacks following an order imposed by other
measures of the centrality of a node, e.g. the so-called betweenness
centrality \cite{Holme02}. In particular for the world-wide airport
network, it has been shown recently \cite{Guimera04,Guimera05} that
nodes with higher betweenness play a more important role in keeping
the network connected than those with high degree. In our study, we
will make use of the scenarios proposed so far as well as develop
further algorithms to perform network attacks.

The paper is organized as follows, in the next Section we describe
the database, define observables in terms of which we are going to
follow the changes in the network properties under attacks, and
describe the different attack strategies we will use. We display our
principal results in sections \ref{III}, \ref{IV}. There, we
formulate criteria that allow to estimate the resilience of networks
against attacks and discuss behavior of the PTNs during attacks
following different strategies, outlining the most effective ones.
Conclusions and an outlook are given in Section \ref{V}.

\section{Databases, observables, and attack strategies}\label{II}

This study continues our analysis of the properties of PTNs
initiated in Refs. \cite{Ferber,Ferber09a,Ferber09b}. As in these
works, we rely on the publicly available information about PTNs of a
set of fourteen major cities of the world \cite{database}. Our
choice for the selection of these cities was motivated by the idea
to collect network samples from cities of different geographical,
cultural, and economical background. In Table \ref{tab1} we give
some information summarizing the empirical analysis of some of the
properties of the PTNs under consideration.

\begin{figure*}[ht]
\tabcolsep5mm
\begin{tabular}{ccc}
\includegraphics[scale=0.9]{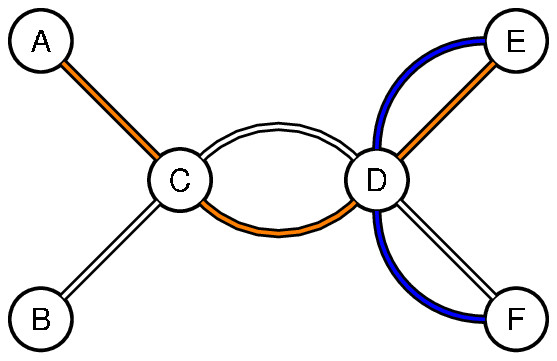} &
\includegraphics[scale=0.9]{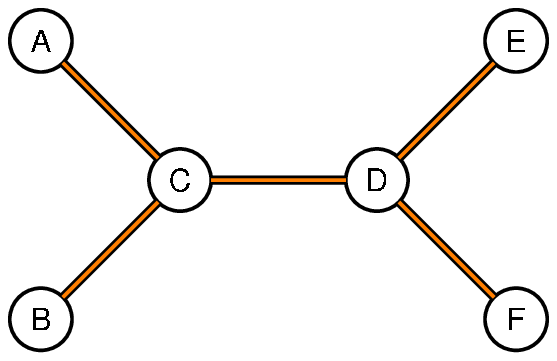} &
\includegraphics[scale=0.9]{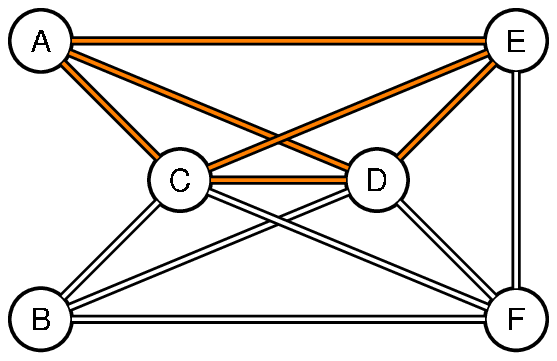} \\  & &
\\
  {\bf  a.}  & {\bf  b.}  &  {\bf  c.}
\end{tabular}
   \caption{\label{fig1}(color online) {\bf a}: a simple public transport map.
    Stations A-F are serviced by routes No 1 (shaded orange), No 2 (white), and No 3 (dark blue).
   {\bf b}: $\mathds{L}$-space graph. {\bf c}: $\mathds{P}$-space graph, the complete sub-graph
   corresponding to route No 1 is highlighted (shaded orange).
   }
\end{figure*}

\begin{table*}[th]
\begin{center}
\tabcolsep1.2mm
 {\small
\begin{tabular}{lrrrrrrrrrrrrrrrrl}
\hline\hline City & Type &   $N$ & $R$ & $\langle k_{\mathds{L}}
\rangle$ & $\ell_{\mathds{L}}^{\rm max}$ & $\langle
\ell_{\mathds{L}} \rangle$ & $c_{\mathds{L}}$ &
$\kappa^{(z)}_{\mathds{L}}$ & $\kappa^{(k)}_{\mathds{L}}$ &
$\gamma_{\mathds{L}}$ & $\langle k_{\mathds{P}} \rangle$    &
$\ell_{\mathds{P}}^{\rm max}$ & $\langle \ell_{\mathds{P}} \rangle$
& $c_{\mathds{P}}$ &
$\kappa^{(z)}_{\mathds{P}}$   &  $\kappa^{(k)}_{\mathds{P}}$   & $\gamma_{\mathds{P}}$  \\
\hline
Berlin       &  BSTU  &  2992  &   211 &  2.58  &  68  &  18.5 &  52.8   & 1.96 & 3.16 & (4.30) &  56.61 & 5  & 2.9 &  41.9 & 11.47 &  84.51 & (5.85)  \\
Dallas       &  B     &  5366  &   117 &  2.18  & 156  &  52.0 &  55.0   & 1.28 & 2.35 & 5.49   & 100.58 & 8  & 3.2 &  48.6 & 11.23 & 145.65 & (4.67)  \\
D\"usseldorf &  BST   &  1494  &   124 &  2.57  &  48  &  12.5 &  24.4   & 1.96 & 3.16 & 3.76   &  59.01 & 5  & 2.6 &  19.7 & 10.56 &  91.17 & (4.62)  \\
Hamburg      &  BFSTU &  8084  &   708 &  2.65  & 156  &  39.7 &  254.7  & 1.85 & 3.26 & (4.74) &  50.38 & 11 & 4.7 & 132.2 &  7.96 &  79.43 & 4.38    \\
Hong Kong    &  B     &  2024  &   321 &  3.59  &  60  &  11.0 &  60.3   & 3.24 & 5.34 & (2.99) & 125.67 & 4  & 2.2 &  11.7 & 10.20 & 232.73 & (4.40)  \\
Istanbul     &  BST   &  4043  &   414 &  2.30  & 131  &  29.7 &  41.0   & 1.54 & 2.69 & 4.04   &  76.88 & 6  & 3.1 &  41.5 & 10.59 & 140.13 & (2.70)  \\
London       &  BST   &  10937 &   922 &  2.60  &  107 &  26.5 &  320.6  & 1.87 & 3.22 & 4.48   &  90.60 & 6  & 3.3 & 90.0  & 16.94 & 166.95 & 3.89    \\
Los Angeles  &  B     &  44629 &  1881 &  2.37  &  210 &  37.1 & 645.3   & 1.59 & 2.73 & 4.85   & 97.99  & 11 & 4.4 & 399.6 & 17.21 & 159.86 & 3.92    \\
Moscow       &  BEST  &  3569  &   679 &  3.32  &  27  &   7.0 &  127.4  & 6.25 & 7.91 & (3.22) &  65.47 & 5  & 2.5 &  38.0 & 26.48 & 130.65 & (2.91)  \\
Paris        &  BS    &  3728  &   251 &  3.73  &  28  &   6.4 &  78.5   & 5.32 & 6.93 & 2.62   &  50.92 & 5  & 2.7 &  59.6 & 24.06 & 88.89  & 3.70    \\
Rome         &  BT    &  3961  &   681 &  2.95  &  87  &  26.4 &  163.4  & 2.02 & 3.67 & (3.95) &  69.05 & 6  & 3.1 &  41.4 & 11.34 & 108.08 & (5.02)  \\
Sa\~o Paolo  &  B     &  7215  &   997 &  3.21  &  33  &  10.3 &  268.0  & 4.17 & 5.95 & 2.72   & 137.46 & 5  & 2.7 &  38.2 & 19.61 & 333.73 & (4.06)  \\
Sydney       &  B     &  1978  &   596 &  3.33  &  34  &  12.3 &  82.9   & 2.54 & 4.37 & (4.03) &  42.88 & 7  & 3.0 &  33.6 & 7.79  & 74.63  & (5.66)  \\
Taipei       &  B     &  5311  &   389 &  3.12  &  74  &  20.9 &  186.2  & 2.42 & 4.02 & (3.74) & 236.65 & 6  & 2.4 &  15.4 & 12.96 & 415.46 & (5.16)  \\
\hline \hline
\end{tabular}
}
\end{center}
\caption{Some characteristics of the PTNs analyzed in this study.
Types of transport taken into account: {\underline B}us, {\underline
E}lectric trolleybus, {\underline F}erry, {\underline S}ubway,
{\underline T}ram, {\underline U}rban train; $N$: number of
stations; $R$: number of routes.  The following characteristics are
given in $\mathds{L}$- and $\mathds{P}$-spaces, as indicated by the
subscripts:  $\langle k \rangle$ (mean node degree); $\ell^{\rm
max}$, $\langle \ell \rangle$ (maximal and mean shortest path
length); $c$ (relation of the mean clustering coefficient to that of
the classical random graph of equal size); $\kappa^{(z)}$,
$\kappa^{(k)}$ (c.f. Eqs. (\ref{1}), (\ref{12})); $\gamma$ (an
exponent in the power law (\ref{2}) fit, bracketed values indicate
less reliable fits, see text). More data is given in
\cite{Ferber09a}. \label{tab1}}
\end{table*}

There are various ways to represent a PTN in terms of a graph
\cite{ptns}. These different representations allow for a
comprehensive analysis of various PTN properties reflecting their
operating functions. It is natural to perform the analysis of PTN
attack resilience in terms of these representations. These are
briefly summarized in Fig. \ref{fig1}. For the purpose of the
present analysis, we will make use of the so-called $\mathds{L}$ and
$\mathds{P}$-space graphs. In $\mathds{L}$-space representation
\cite{ptns} the PTN is represented by a graph with nodes that
correspond to the stations, whereas links correspond to connections
between stations within one stop distance (Fig. \ref{fig1}b). In the
$\mathds{P}$-space \cite{Sen03} all station-nodes that belong to the
same route form of a complete subgraph of the network (Fig.
\ref{fig1}c).

\begin{figure*}[ht]
\centerline{\includegraphics[width=80mm]{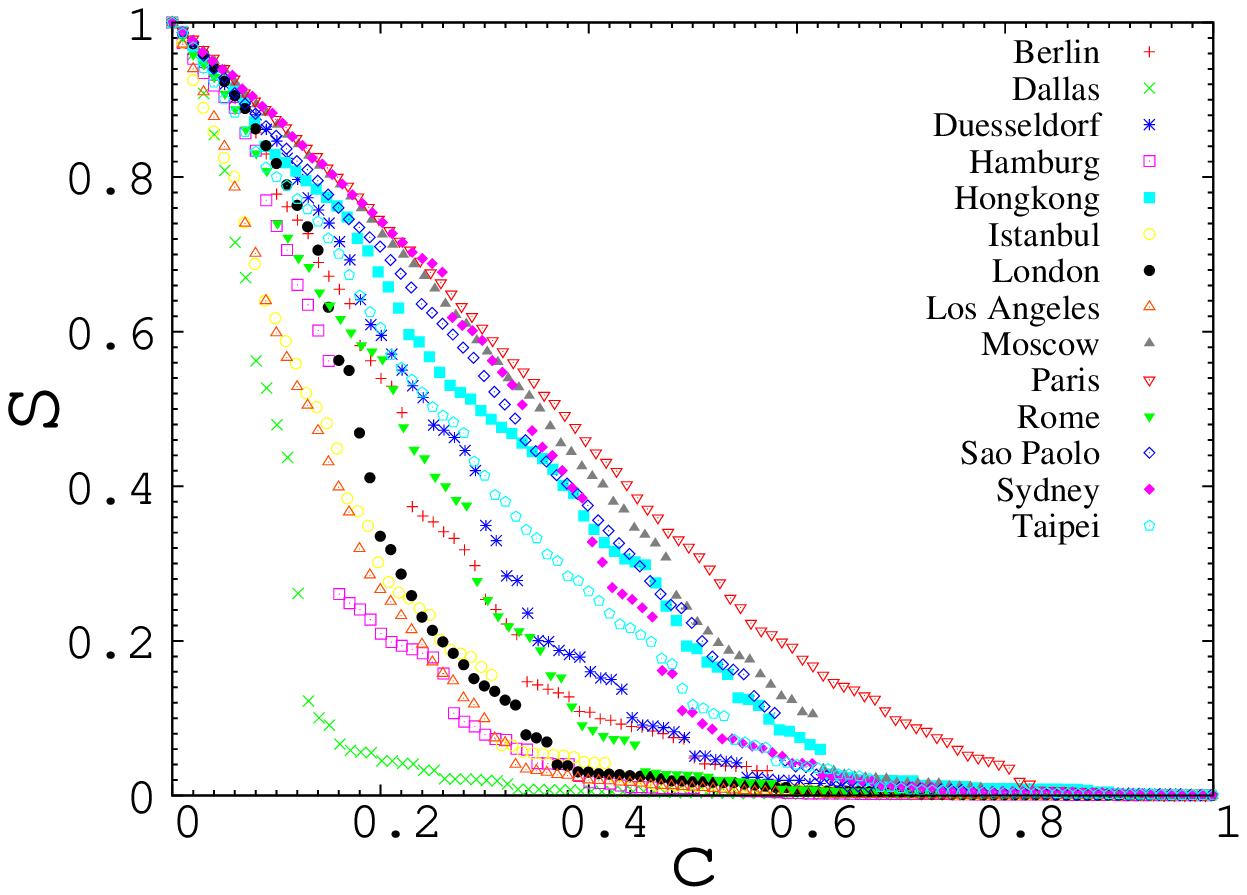}
\hspace*{3em}
\includegraphics[width=80mm]{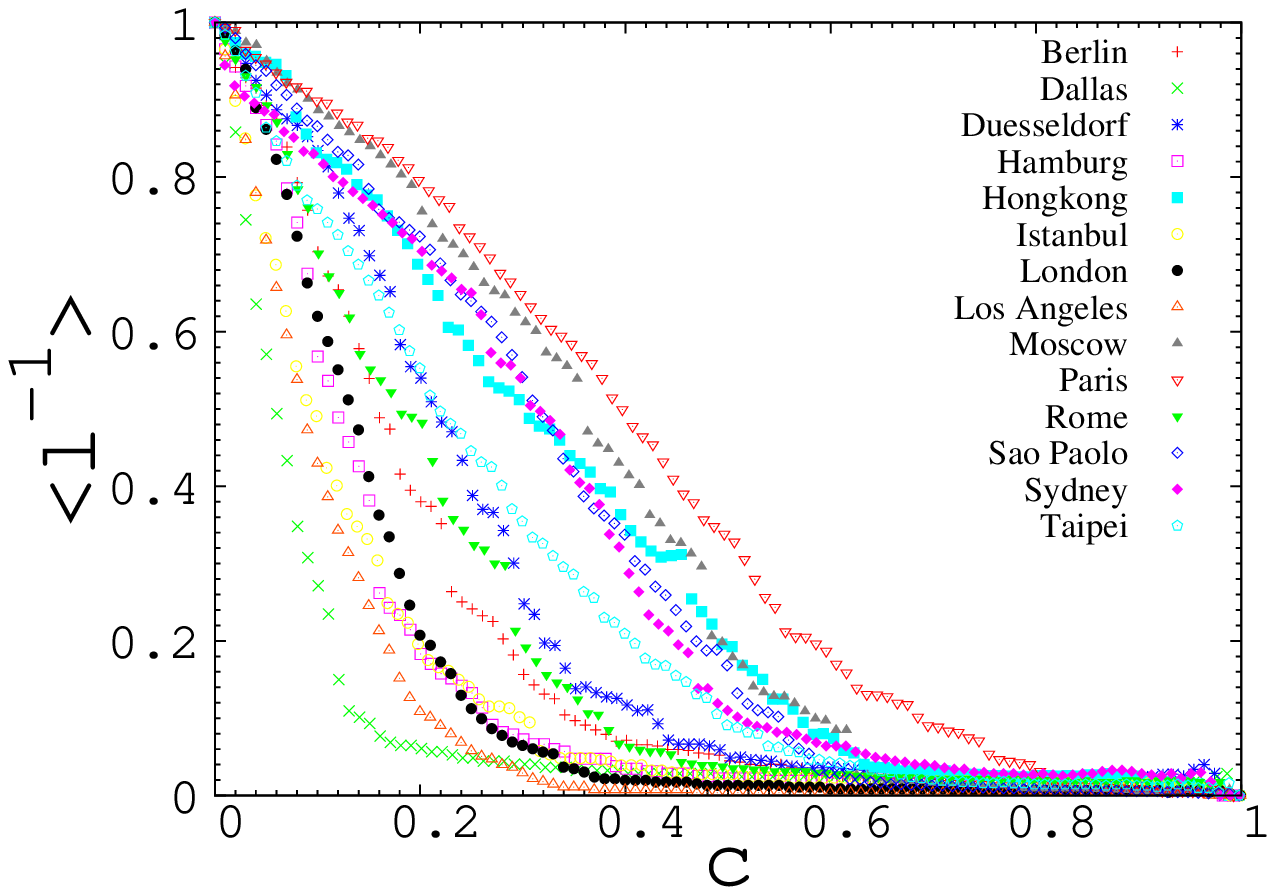}}
\vspace*{3ex} \centerline{{\bf a.} \hspace*{30em} {\bf b.}}
 \caption{(color online). $\mathds{L}$-space. Random scenario. Size of the largest cluster $S$ ({\bf a.})
 and an average inverse
mean shortest path length $\langle\ell^{-1}\rangle$ ({\bf b.})
 as functions of a fraction of
removed nodes $c$ normalized by their values at $c=0$. \label{fig2}}
\end{figure*}

Let us take the $\mathds{L}$-space representation to introduce the
observables we will use to quantify the PTN behavior under attack.
Keep in mind however, that in our analysis presented in Section
\ref{III} we will deal also with the $\mathds{P}$-space. There are
two intrinsically connected questions that naturally arise when one
wants to describe quantitatively how a certain network changes when
its nodes are removed \cite{note3}. The first is how to choose the
'order-parameter' variable that signals the quantitative change in
the network behavior (i.e. the break down of the network), the
second is how to locate the value of concentration of removed nodes
at which this change occurs. As we have mentioned in the
introduction, in a theoretical description a useful quantity is the
GCC: its disappearance can be associated with a network breakdown.
Strictly speaking, the GCC is well-defined only in the $N\to \infty$
limit, therefore in practice dealing with a network of a finite size
$N$ it is substituted by the size of the largest connected
component. We will use in the following its normalized value defined
by:
\begin{equation} \label{3}
S=N_1/N,
\end{equation}
with $N$ and $N_1$ being number of nodes of the network and of its
largest component correspondingly. By definition (\ref{3}), a
largest component is always present in a network of non-zero size. A
useful quantity to measure network connectivity is the average
shortest path:
\begin{equation}\label{4}
\langle \ell \rangle = \frac{2}{N(N-1)}\sum_{i>j}\ell (i,j),
\end{equation}
where $\ell (i,j)$ is the length of a shortest path from node $i$ to
$j$ and the sum spans all pairs $i,j$ of sites of the network.
However, $\langle \ell \rangle$ is ill-defined for a disconnected
network. Alternatively, one can suitably define the mean inverse
shortest path length \cite{Holme02} by:
\begin{equation}\label{5}
\langle \ell^{-1} \rangle = \frac{2}{N(N-1)}\sum_{i>j}\ell^{-1}
(i,j),
\end{equation}
with $\ell^{-1} (i,j)=0$ if nodes $i,j$ are disconnected. As one can
see, Eq. (\ref{5}) is well-defined even for a disconnected network
and as such can be used to trace changes of network behavior  under
attack. To give an example, we show in Fig. \ref{fig2} how the
largest component fraction $S$, Eq. (\ref{3}) and the mean inverse
shortest path length $\langle \ell^{-1} \rangle$, Eq. (\ref{5}),
change upon random removal of nodes in each of fourteen PTNs
selected for our study. More precisely, we measure these quantities
as functions of the fraction of removed nodes $c$ starting from the
unperturbed network ($c=0$) and eliminating at random step-by-step 1
\% of the nodes up to $c=1$. In what follows below we will call this
scenario a {\em random scenario}.

Already this first attack attempt brings about interesting (and in
part unexpected) PTN features. Namely:

\noindent (i) different PTNs react on random removal of their nodes
in different ways, that range from rapid abrupt breakdown (Dallas)
to a slow almost linear decrease (Paris);

\noindent (ii) although qualitatively similar, the observed impact
of the attack differs depending on which variable is used as
indicator, either $S$ or $\langle \ell^{-1} \rangle$. Ordering the
PTNs by their vulnerability, this order may thus differ depending on
the applied indicator;

\noindent (iii)  up to $c=1$, there is no general 'percolation
threshold' concentration of removed nodes $c$ at which $S$ (or
$\langle \ell^{-1} \rangle$) vanishes that would hold for all PTNs.
Rather for some individual PTNs one observes various values of $c$
at which these PTNs show abrupt changes of their properties.

Figs. \ref{fig2} {\bf a},{\bf b} display how the different PTNs
react on a {\em random} removal of their nodes. Obviously, the
question immediately arises how this behavior changes if one removes
the nodes not at random, but following a given order or scheme (we
call this the scenario of the attack). As we have mentioned in the
introduction, a number of different attack scenarios have been
proposed
\cite{Albert00,Callaway00,Holme02,Barabasi99,Broder00,Girvan02,%
Guimera04,Guimera05,Ferber09b}. These are generally based on the
intuitive assumption  that the largest impact on a network is caused
by the removal of its most 'important' nodes. A number of indicators
have been developed in particular in applications of graph theory
for social science to measure the importance of a node. Besides the
node degree $k_j$, which is equivalent to the number of nearest
neighbors $z_1(j)$ of a given node $j$, different centralities have
been introduced for this purpose. In particular, the closeness
$C_C(j)$, graph $C_G(j)$, stress $C_S(j)$, and betweenness
centralities $C_B(j)$ of a node $j$ are defined as follows (see e.g.
\cite{Brandes01}):
\begin{eqnarray}\label{6}
 C_C(j) &=& \frac{1}{\sum_{t\in \cal{N}} \ell(j,t)}, \\
\label{7}
 C_G(j) &=& \frac{1}{{\rm
max}_{t\in \cal{N}} \ell(j,t)}, \\
\label{8}
 C_S(j) &=& \sum_{s\neq j \neq t\in \cal{N}} \sigma_{st}(j), \\
\label{9}
 C_B(j) &=& \sum_{s\neq j \neq t\in \cal{N}}
 \frac{\sigma_{st}(j)}{\sigma_{st}}.
 \end{eqnarray}
In Eqs. (\ref{6})--(\ref{8}), $\ell(j,t)$ is the length of a
shortest path between the nodes $j,t$ that belong to the network
$\cal{N}$, $\sigma_{st}$ is the number of shortest paths between the
two nodes $s,t\in \cal{N}$, and $\sigma_{st}(j)$ is the number of
shortest paths between nodes $s$ and $t$ that go through the node
$j$. Alternatively, one may measure the importance of a given node
$j$ by the number of its second nearest neighbors $z_2(j)$ or its
clustering coefficient $C(j)$. The latter is the ratio of the number
of links $E_j$ between the $k_j$ nearest neighbors of $j$ and the
maximal possible number of mutual links between them:
\begin{equation} \label{10}
C(j)  =\frac{2E_j}{k_j(k_j-1)}.
\end{equation}

\begin{figure}
\centering
\includegraphics[width=80mm]{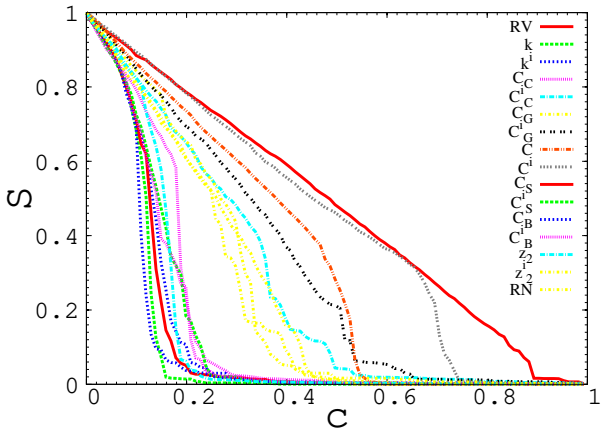}
 \caption{(color online). Largest component size of the PTN of Paris as function of  the fraction of
 removed nodes for different attack scenarios. Each curve corresponds to a different scenario as indicated
in the legend. Lists of removed nodes were prepared according to
their degree $k$, closeness $C_C$, graph $C_G$, stress $C_S$, and
betweenness $C_B$ centralities, clustering coefficient $C$, and next
nearest neighbors number $z_2$. A superscript $i$ refers to lists
prepared for the initial PTN before the attack. RV and RN denote the
removal of a random vertex (RV) or of its randomly chosen neighbor
(RN), respectively.}
 \label{fig3}
\end{figure}

Removing important nodes according to lists prepared in the order of
decreasing node degrees $k$, centralities (\ref{6})--(\ref{9}),
number of their second nearest neighbors $z_2$, and increasing
clustering coefficient $C$ defines seven different attack scenarios.
As we have already mentioned in the introduction, the scenarios can
be either implemented according to lists prepared for the {\em
initial} PTN before the attacks (we will indicate the corresponding
scenario by a superscript $i$, e.g. $C_B^{i}$) or by lists rebuilt
by recalculating the order of the remaining nodes after each step.
Together, this leads to fourteen different attack scenarios. In
addition, we will keep the above described random scenario (denoted
further as RV) and add one scenario more, removing a randomly chosen
neighbor of a randomly chosen node (RN). The latter scenario appears
to be effective for immunization problems \cite{Cohen03} and it is
based on the fact, that in this way nodes with a high number of
neighbors will be selected with higher probability. Note that in
this scenario only a neighbor node is removed and not the initially
chosen one.

All together, this defines sixteen different scenarios to attack a
network and we apply these to all fourteen PTNs that form our
database. A typical result for a single PTN is displayed in Fig.
\ref{fig3}. Here, we show how the largest connected component size
$S$ of the Paris PTN changes under the influence of the above
described attack scenarios. Already from this plot one may
discriminate between the most effective scenarios that result in a
fast decrease of the largest component size (those governed by
betweenness and stress centralities, node degree, and next nearest
neighbors number -- see the Figure) and the less harmful ones. In
the following, instead of displaying the results of all attacks for
all different PTNs we will focus on the results of the most
effective scenarios comparing them with those of random failure as
introduced by the random scenario. As outlined in the introduction,
we make use of different PTN representations (different 'spaces' of
Fig. \ref{fig1}). In the following section, we present the analysis
of PTN resilience in the $\mathds{L}$-space representation.

\begin{figure*}[htb]
\centerline{\includegraphics[width=80mm]{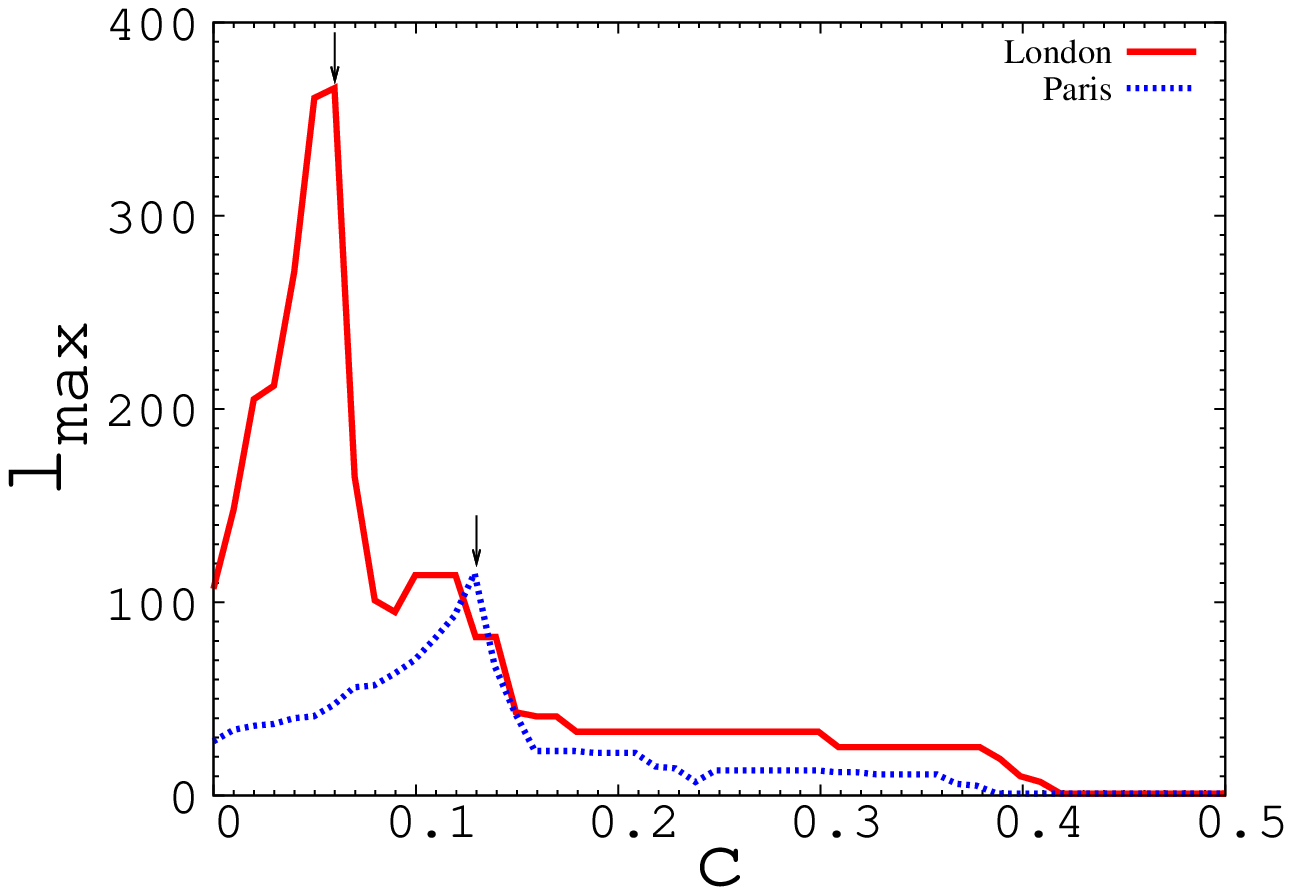}
\hspace*{3em}
\includegraphics[width=80mm]{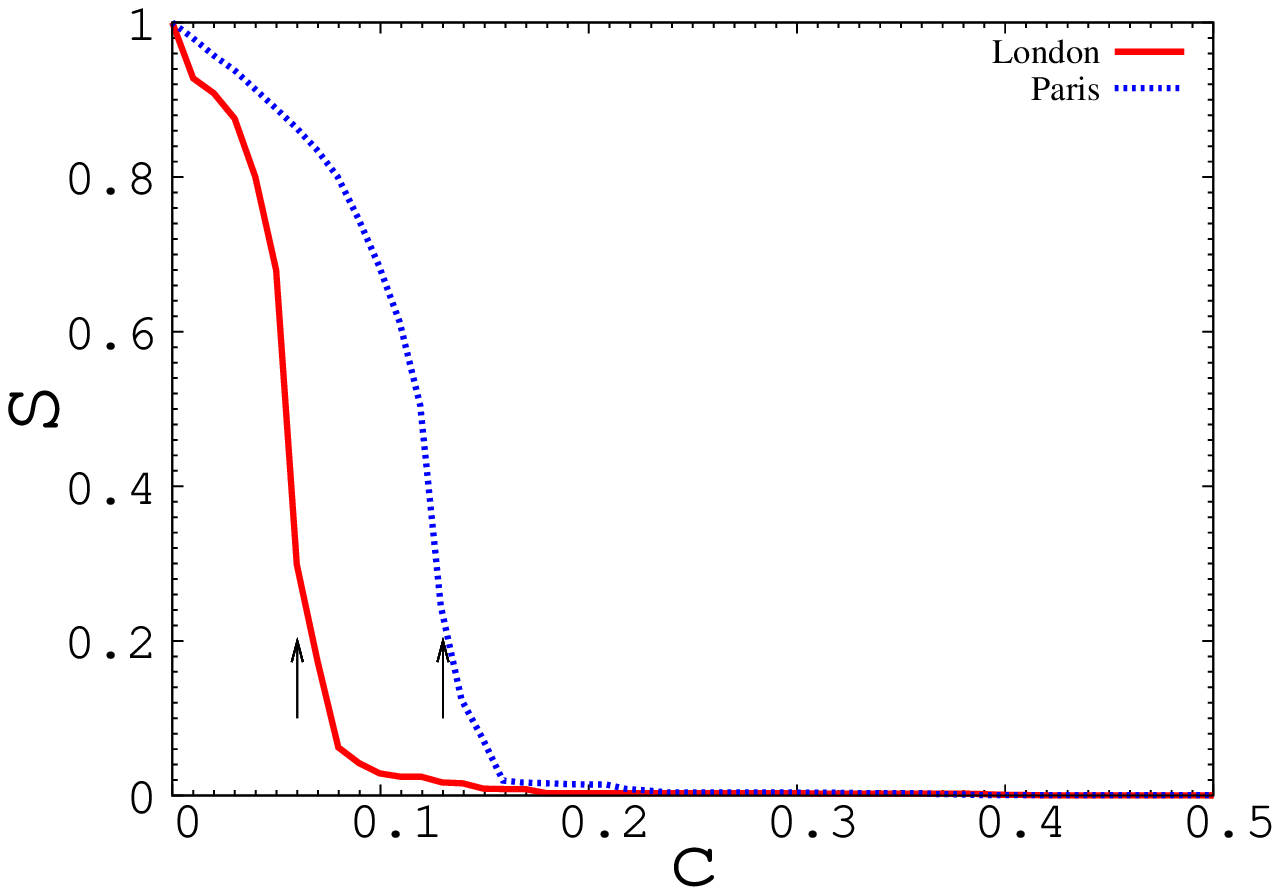}}
\vspace*{3ex} \centerline{{\bf a.} \hspace*{30em} {\bf b.}}
\caption{(color online). $\mathds{L}$-space. Recalculated highest
degree scenario. {\bf a.} behavior of the maximal shortest path
$\ell_{\rm max}$ for the PTNs of Paris and London. Note the
characteristic peaks that occur at $c=0.13$ (Paris) and $c=0.06$
(London). {\bf b.} Size of largest connected cluster $S$ as function
of a fraction of removed nodes for the same networks. The arrows
indicate the values of $c$ at which the peak for $\ell_{\rm max}$
appears. \label{fig4}}
\end{figure*}

\section{Results in $\mathds{L}$-space}\label{III}

The $\mathds{L}$-space representation of a PTN is a graph that
represents each station by a node, a link between nodes indicates
that there is at least one route that services the two corresponding
stations consecutively. No multiple links are allowed (see Fig.
{\ref{fig1}\bf b}). Therefore, attacks in the $\mathds{L}$-space
correspond to situations, in which given public transport stations
cease to operate for all means of traffic that go through them. Note
however, that in this representation, the removal of a station node
does not otherwise interfere with the operation of a route that
includes this station. It rather splits this route into two
(operating) pieces. An alternative situation will be considered in
the forthcoming section.

In order to answer some of the questions raised in Section \ref{II},
let us return to Fig. \ref{fig3}, where the impact on the largest
component size $S$ of the PTN of Paris is shown for sixteen
different attack scenarios as function of the fraction of removed
nodes. As we have already remarked, for this PTN the most
influential are the scenarios where nodes are removed according to
lists ordered by $C_B$, $k$, $C_S$, $k^{i}$, $C_B^{i}$, $C_S^{i}$
(we list the characteristics in a decreasing order of effectiveness
of the corresponding scenario). For a small value of $c$ ($c<0.07$)
these scenarios cause practically indistinguishable impact on $S$
with a linear behavior $S\sim (1-c)$. As $c$ increases, deviations
from the linear behavior arise and the impact of different scenarios
start to vary. In particular, there appear differences between the
role played by the nodes with highest value of $k$ and highest
betweenness centrality $C_B$. Whereas the first quantity is a local
one, i.e. it is calculated from properties of the immediate
environment of each node, the second one is global. Moreover, the
$k$-based strategy aims to remove a maximal number of edges whereas
the $C_B$-based strategy aims to cut as many shortest paths as
possible. In addition, there arise differences between the 'initial'
and 'recalculated' scenarios, suggesting that the network structure
changes as important nodes are removed. Similar behavior of $S(c)$
is observed for all PTNs included in this study, with certain
peculiarities in the order of effectiveness of different attack
scenarios. Note however, that the difference between 'initial' and
'recalculated' scenarios is less evident for strategies based on
local characteristics, as e.g. the node degree or the number of
second nearest neighbors (c.f. curves for $k$, $k^i$ and $z_2$,
$z_2^i$, respectively). This difference between initial and
recalculated characteristics is even more pronounced for the
centrality-based scenarios.

Now let us return to some of the observations of Section \ref{II}.
Namely, we noted that the observed impact of an attack may differ
depending on which observable is used as the 'order-parameter'
variable (c.f. Fig. \ref{fig2} where this is shown for the RV attack
scenario taking either $S$ or $\langle \ell^{-1} \rangle$ as
'order-parameter'). Similar differences we observe also in the case
of the other scenarios. For the sake of uniqueness in the following
we will use the value of $S$ to measure the effectiveness of a given
attack. This choice is motivated by several reasons: (i) in an
infinite network limit $S$ defines an order parameter of the
classical percolation problem \cite{percolation}; (ii) differences
between network resilience as judged e.g. by the behavior of $S$ or
by that of $\langle \ell^{-1} \rangle$ are not significant enough to
be a subject of special analysis (at least not for the PTNs we
consider); (iii) considering $S$ naturally leads to other useful
characteristics that allow to estimate the PTN operating ability and
its segmentation. Let us stop to elaborate the latter point in more
detail.

As we have already emphasized, there is no well defined 'percolation
threshold' concentration of removed nodes $c_{\rm perc}$ at which
$S$ (or $\langle \ell^{-1} \rangle$) vanishes (see Figs. \ref{fig2},
\ref{fig3}) which could serve as  evidence of a break down of the
largest PTN component and hence of the loss of operating ability
\cite{note4}. In Ref. \cite{Ferber09b} it has been proposed to use
the behavior of maximal shortest path length $\ell_{\rm max}$ as a
possible indicator of the network break down. This was based on the
observation, that as the concentration of removed nodes $c$
increases, the value of $\ell_{\rm max}$ for different PTNs displays
similar typical behavior: initial growth and then an abrupt decrease
when a certain threshold is reached (see e.g. Fig. \ref{fig4} {\bf
a} where this value is shown for the recalculated highest degree
attack scenario of the PTNs of Paris and London). Obviously,
removing the nodes initially increases the path lengths as
deviations from the original shortest paths need to be taken into
account. Further removing nodes then at some point leads to the
breakup of the network into smaller components on which the paths
are naturally limited by the size of these components which explains
the sudden decrease of their lengths. For comparison, in Fig.
\ref{fig4} {\bf b} we show how the value of $S$ changes under the
recalculated highest degree scenario for the above PTNs.

Being certainly useful for many instances of the PTNs analyzed, the
above $\ell_{\rm max}$-based criterion cannot serve as an universal
tool to determine the region of $c$, where the network stops to
operate. One of the reasons is that for certain PTNs (as well as for
certain attack scenarios) we have found that $\ell_{\rm max}$ does
not show a pronounced maximum, but rather shows several maxima at
different values of $c$. Therefore, to devise a criterion which may
be equally well used for any of the networks we decided to define
characteristic concentration of removed nodes $c_{\rm s}$ at which
the size of the largest component $S$ decreases to one half of its
initial value. This characteristic concentration allows us to
compare the effective robustness of different PTNs or of the same
PTN when different attack scenarios are applied. In what follows
below, we will call this concentration the {\em segmentation}
concentration $c_{\rm s}$, with the obvious condition:
\begin{equation} \label{11}
S(c_{\rm s})  = \frac{1}{2}S(c=0).
\end{equation}

\begin{figure}
\centering
\includegraphics[width=80mm]{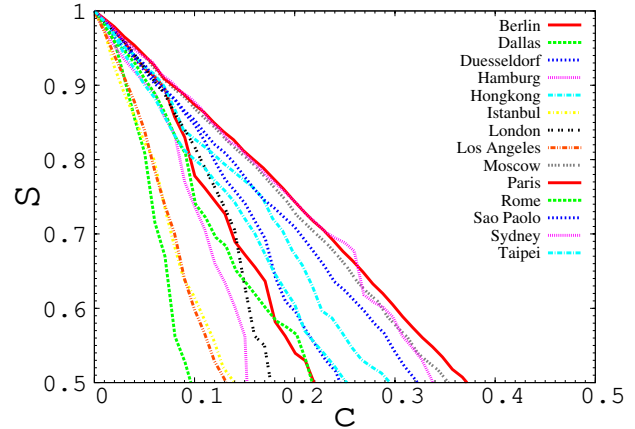}
 \caption{(color online). $\mathds{L}$-space. Random scenario. Size of the largest cluster $S$
normalized by its value at $c=0$ as function of a fraction of
removed nodes. From this figure it is easy to define the fraction of
nodes $c_{\rm s}$ which satisfies Eq. (\ref{11}).}
 \label{fig5}
\end{figure}

In Fig. \ref{fig5} we plot the size of the largest connected
component $S$ for different PTNs as function of the fraction of
removed nodes $c$  for the random vertex scenario (RV) in
$\mathds{L}$-space. The choice of the lowest $S$ value $S=1/2$ in
this figure enables one to find the value $c_{\rm s}$ as the
crossing point of $S(c)$ with the horizontal axis. The values of
$c_{\rm s}$ obtained for this scenario are given in the last column
of Table \ref{tab2}. Note that the PTNs under consideration react on
random attack in many different ways: some of them slowly decrease
without any abrupt changes in $S$ (like PTNs of Paris, Moscow,
Sydney) while others are characterized by rather fast decay of $S$
(Dallas, Los Angeles, Istanbul).

\begin{table*}[htbp]
\centering \tabcolsep2.5mm
\begin{tabular}{lllllllllllll}
\hline \hline
           City& $c_{\rm s}$&        & $c_{\rm s}$   &  & $c_{\rm s}$  & & $c_{\rm s}$ &  & $c_{\rm s}$ &  & $c_{\rm s}$ &  \\ \hline
 Berlin        &     .060   & $C_B$  & .065 & $k^{i}$& .065 & $C_S$  & .070 & $k$   & .075& $z_2$    &  .220 & RV    \\
 Dallas        &     .025   & $k^{i}$& .030 & $k$    & .030 & $C_B$  & .045 & $z_2$ & .055& $z_2^{i}$&  .090 & RV  \\
 D\"usseldorf  &     .075   & $C_B$  & .080 & $k$    & .080 & $k^{i}$& .095 & $C_S$ & .105& $z_2$    &  .240 & RV   \\
 Hamburg       &     .040   & $C_B$  & .040 & $C_C$  & .045 & $C_S$  & .045 & $k^{i}$&.060& $z_2$    &  .150 & RV   \\
 Hong Kong     &     .030   & $C_B$  & .040 & $C_C$  & .050 & $z_2^{i}$ & .060 & $C_S$&.090& $k^{i}$ &  .300 & RV   \\
 Istanbul      &     .025   & $C_S$  & .030 & $C_C$  & .030 & $C_B$  & .035 & $k^{i}$ &.035& $k$     &  .140 & RV   \\
 London        &     .055   & $k$    & .060 & $k^{i}$& .065 & $C_B$  & .075 & $C_C$ & .085& $z_2$    &  .175 & RV   \\
 Los Angeles   &     .040   & $k$    & .060 & $k^{i}$& .065 & $z_2$  & .075 & $C_B$ & .100& $z_2^{i}$&  .130 & RV   \\
 Moscow        &     .070   & $C_B$  & .085 & $C_S$  & .085 & $k$    & .085 & $k^{i}$&.100& $C_C$    &  .350 & RV   \\
 Paris         &     .105   & $C_B$  & .120 & $k$    & .125 & $C_S$  & .130 & $k^{i}$&.140& $C_B^{i}$&  .375 & RV   \\
 Rome          &     .050   & $C_B$  & .060 & $C_C$  & .065 & $k$    & .065 & $k^{i}$&.085& $C_S$    &  .215 & RV   \\
 Sa\~o Paolo   &     .040   & $k$    & .040 & $k^{i}$& .045 & $C_B$  & .060 & $C_S$ & .060& $C_S^{i}$&  .320 & RV   \\
 Sydney        &     .040   & $C_B$  & .040 & $C_C$  & .065 & $C_S$  & .075 & $k^{i}$&.085& $C_G,k$  &  .350 & RV   \\
 Taipei        &     .105   & $C_B$  & .105 & $C_G$  & .115 & $k$    & .120 &  $k^{i}$&.120&  $C_C$  &  .240 & RV
 \\
 \hline \hline
\end{tabular}
\caption{Segmentation concentration $c_{\rm s}$ for different attack
scenarios applied to different PTNs.  For each city, the Table
displays the results of the five most destructive attack scenarios
ordered by increasing values of $c_{\rm s}$. The scenario is
indicated after corresponding value of $c_{\rm s}$. The scenarios
are abbreviated by the name of the characteristics used to prepare
the lists of removed nodes (see Sec. \ref{II} for detailed
explanation). In the last column the value of $c_{\rm s}$ for the
random scenario (RV) is shown. \label{tab2} }
\end{table*}

Now, applying these attacks according to the sixteen scenarios
described above we are in the position to discriminate them by their
degree of destruction and to single out those with the highest
impact on each of the PTNs considered. To this end, for each PTN we
give in Table \ref{tab2} the segmentation concentration $c_{\rm s}$
for the five most harmful attack scenarios. The obtained values of
$c_{\rm s}$ are given in increasing order. Near each value we denote
the scenario that was implemented. Our analysis reveals the most
harmful scenarios as those targeted at nodes with the highest values
of either the node degree $k$, the betweenness centrality $C_B$, the
next nearest neighbor number $z_2$, or the stress centrality $C_S$
recalculated after each step of the attack.

\begin{figure*}
\includegraphics[width=80mm]{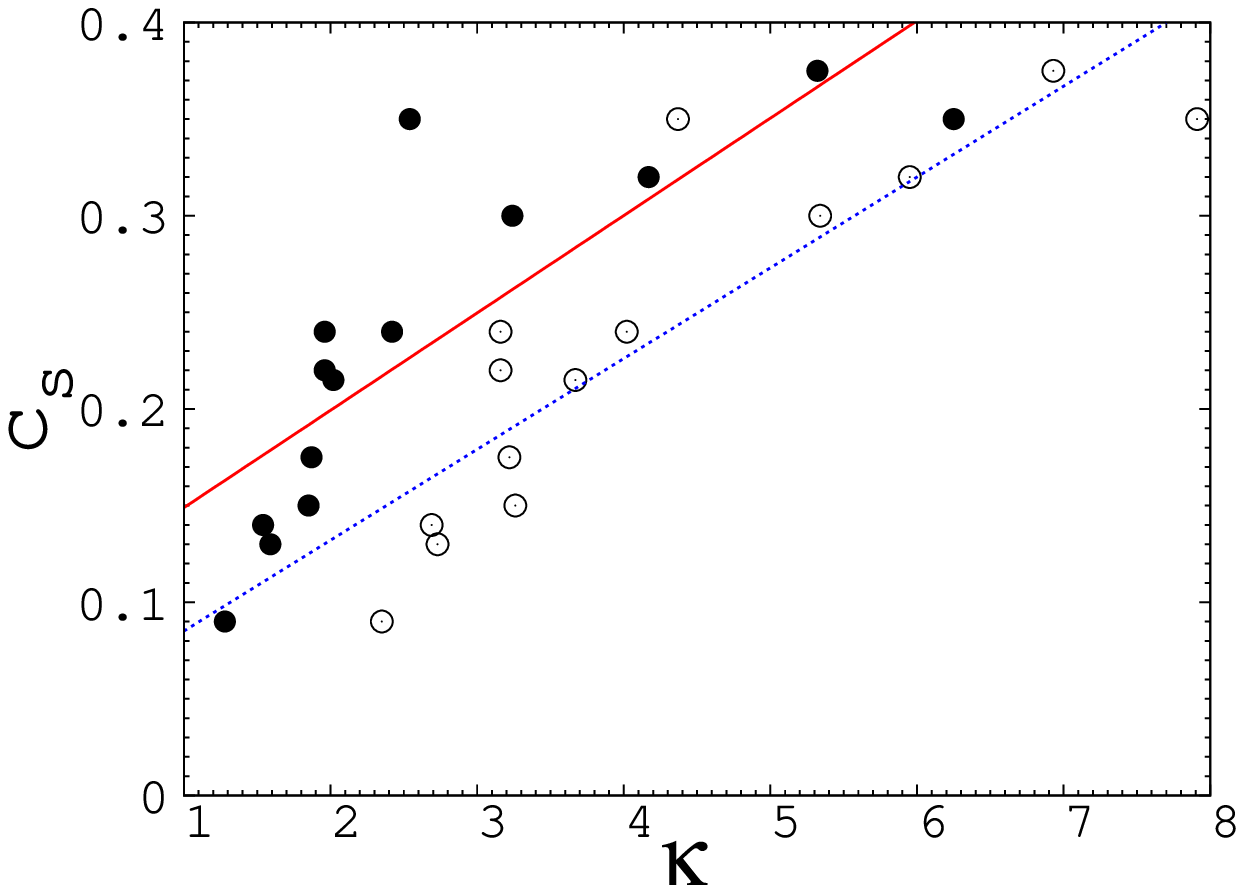}
\hspace*{3em}
\includegraphics[width=80mm]{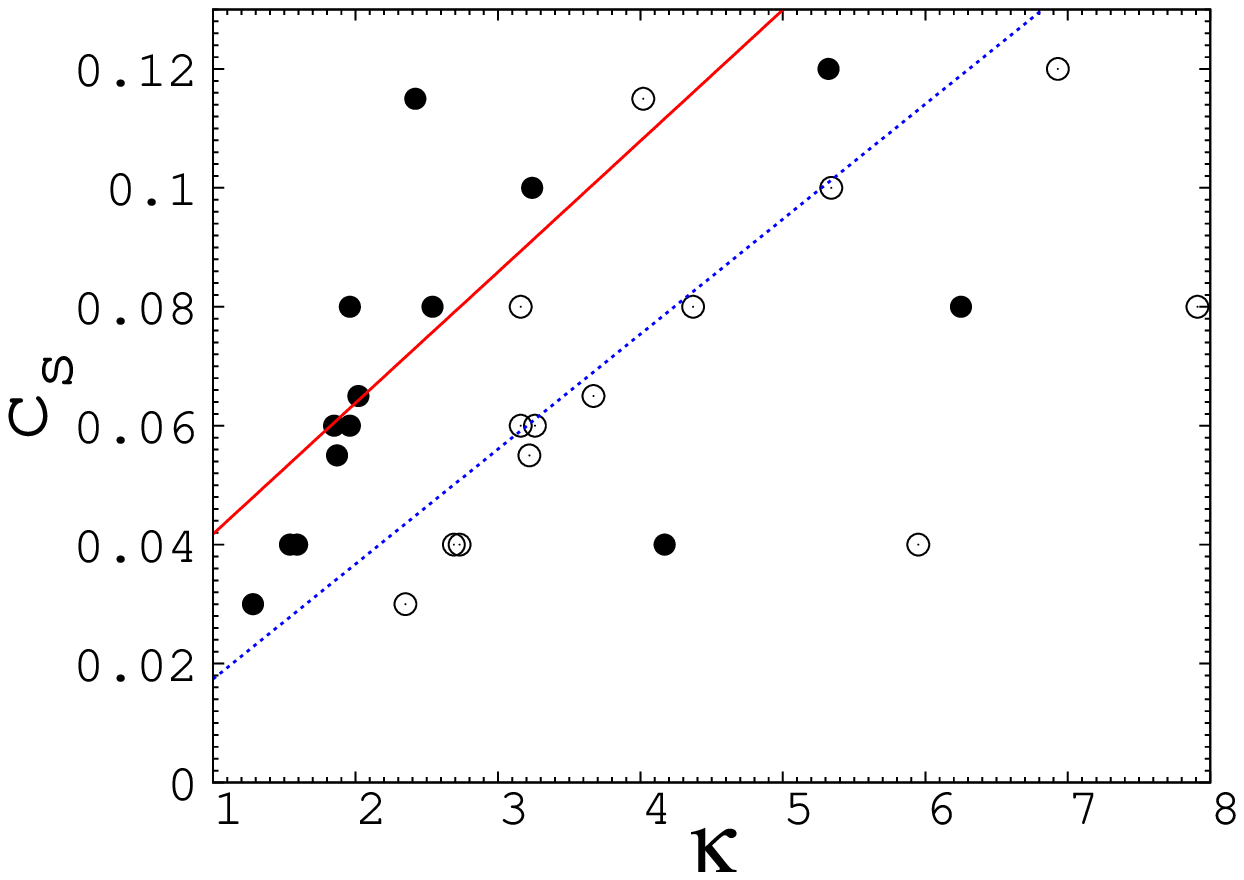}
\vspace*{3ex}
 \caption{$\mathds{L}$-space. Correlations between the ratio $\kappa$, Eq. (\ref{1a}), (\ref{12}) and
 segmentation concentration $c_{s}$. Open circles: $\kappa^{(k)}=\langle k^2\rangle / \langle k \rangle $,
filled circles: $\kappa^{(z)}=z_2/z_1$. The lines serve as guides to
observe the tendency of $c_{\rm s}$ to increase for higher values of
$\kappa$. {\bf a.} {\em Random scenario.} Most out-of-range are the
points $c_{\rm s}=0.35$, $\kappa^{(z)}=2.54$, $\kappa^{(k)}=4.37$
(Sydney)
 and $c_{\rm s}=0.35$, $\kappa^{(z)}=6.25$, $\kappa^{(k)}=7.91$  (Moscow).
 {\bf b.} {\em Recalculated node-degree
 scenario.} Two PTNs are out of range:   $c_{\rm s}=0.04$, $\kappa^{(z)}=4.17$, $\kappa^{(k)}=5.95$ (Sa\~o Paolo)
 and $c_{\rm s}=0.08$, $\kappa^{(z)}=6.25$, $\kappa^{(k)}=7.91$  (Moscow).}
 \label{fig6}
\end{figure*}

It is instructive to observe correlations between the
characteristics of unperturbed PTNs (see Table \ref{tab1}) and their
robustness to attacks. Such correlations may allow for an a priory
estimate of the resilience of a network with respect to attacks. As
discussed in the introduction, percolation  theory for uncorrelated
networks predicts that the value of the Molloy-Reed parameter
$\kappa^{(k)}$, Eq. (\ref{1}), can be used to measure the distance
to the percolation point $\kappa^{(k)}=2$. We may therefore expect
that networks with a higher value of $\kappa^{(k)}$ show higher
resilience. To this end let us first compare the values of $c_{\rm
s}$ for certain scenarios with the value of $\kappa^{(k)}$ for the
unperturbed PTN. Before doing this let us note that for an
uncorrelated  network the value of $\kappa^{(k)}$ can be equally
represented by the ratio between the mean next neighbors number of a
node $z_1$ (which is by definition equal to the mean node degree
$\langle k \rangle$) and the mean second nearest neighbors number
$z_2$:
\begin{equation}\label{12}
\kappa^{(z)}=z_2/z_1.
\end{equation}
Indeed, given that for such a network (see e.g. \cite{networks})
\begin{equation}\label{13}
z_2= \langle k^2 \rangle - \langle k \rangle ,
\end{equation}
one can rewrite (\ref{1a}) as:
\begin{equation}\label{14}
\kappa^{(z)}=1 \hspace{1em}\mbox{ at }\hspace{1em} c_{\rm perc}.
\end{equation}
The relation $\kappa^{(k)}=\kappa^{(z)}+1$ holds only approximately
for the real-world networks we consider in our study, as one can
see, e.g., from the Table \ref{tab1}. In Fig. \ref{fig6}{\bf a} we
compare both quantities $\kappa^{(k)}$, $\kappa^{(z)}$ for
unperturbed PTNs with the corresponding segmentation concentration
$c_{\rm s}$ for the random attack scenario. Within the expected
scatter of data one can definitely observe a general tendency of
$c_{\rm s}$ to increase with both $\kappa^{(k)}$ and $\kappa^{(z)}$:
the higher the value of $\kappa$ for an unperturbed network, the
more robust it is to random removal of its vertices. This
conclusion, however with a more pronounced scatter of data even
holds if one repeats the same analysis for the case of the scenario
based on recalculated node degrees, as shown in Fig. \ref{fig6}{\bf
b}. Again, one observes $c_{\rm s}$ to increase with increasing
$\kappa$. For the betweenness-based attack scenarios the data is
even more scattered and a prediction based on the a priori
calculated ratios is unreliable.

\begin{figure*}
\includegraphics[width=80mm]{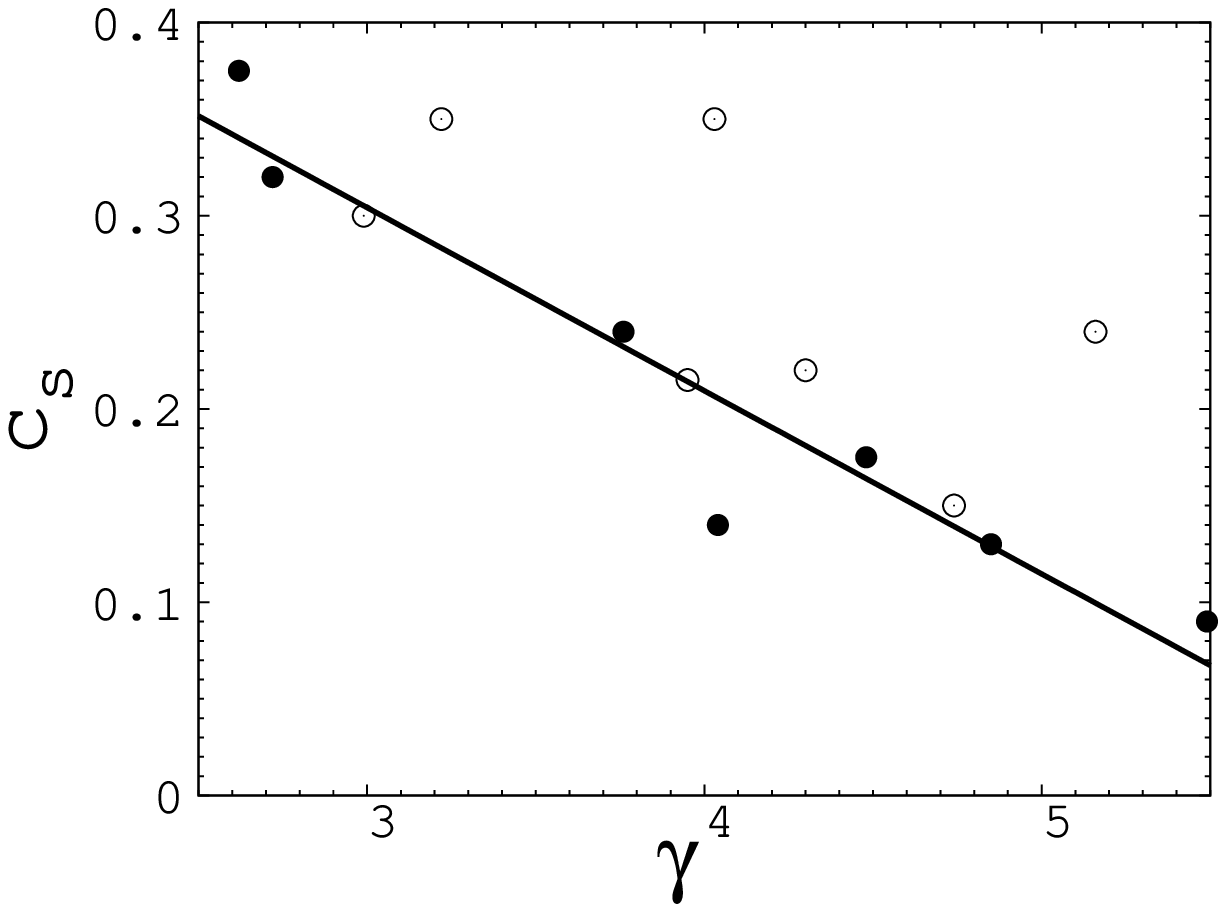} \hspace*{3em}
\includegraphics[width=80mm]{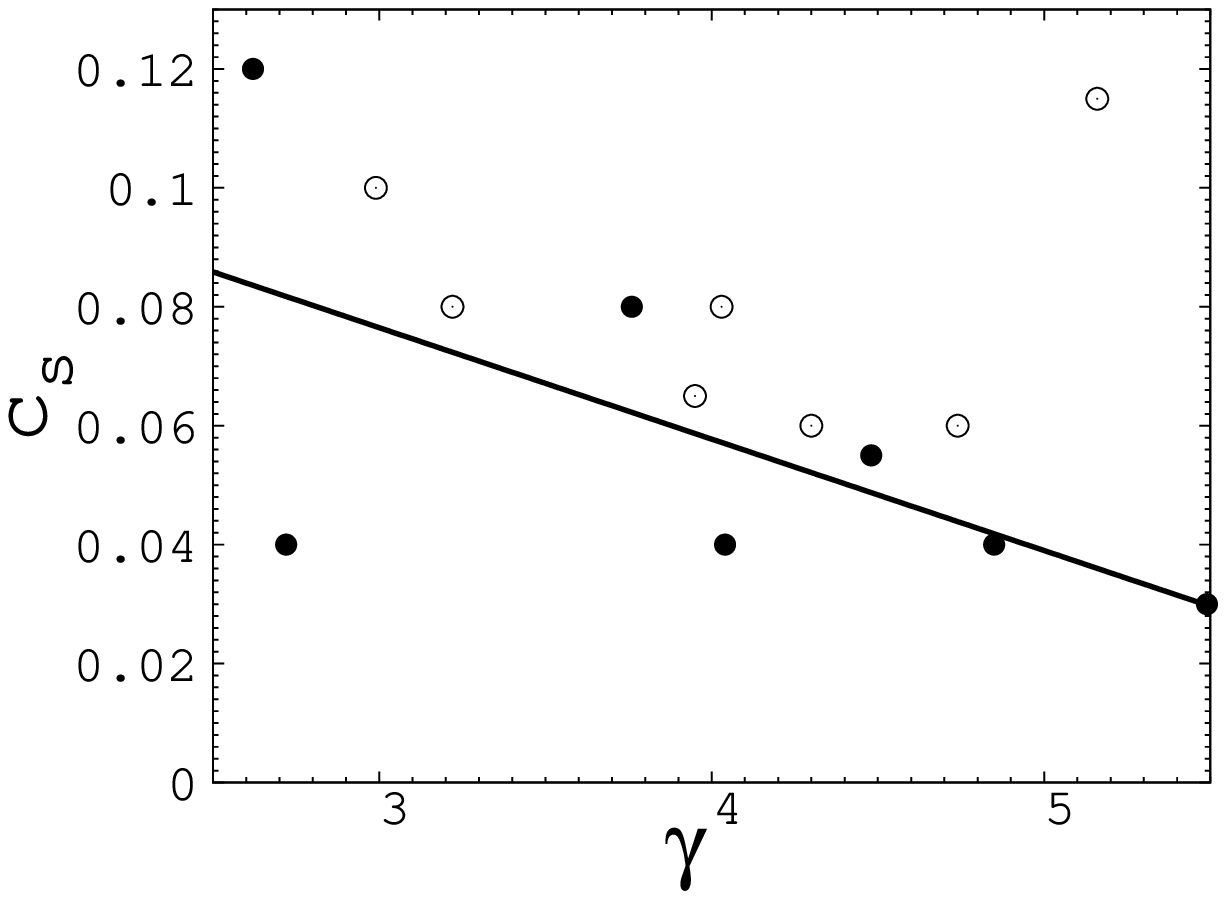}
\vspace*{3ex}
 \centerline{{\bf a.} \hspace*{30em} {\bf b.}}
 \caption{$\mathds{L}$-space. Correlations between the node-degree distribution
 exponent $\gamma$ and segmentation concentration $c_{s}$. Filled circles: scale-free PTNs,
 open circles: PTNs with less pronounced power-law decay. Solid lines serve as guides to observe
 the tendency of $c_{\rm s}$ to decay with an increase of
 $\gamma$. {\bf a}. {\em Random scenario}. Most out of range are the points at
$c_{\rm s}=0.24$, $\gamma=(5.16)$ (Taipei) and at $c_{\rm s}=0.35$,
$\gamma=4.03$ (Sydney). {\bf b}. {\em  Recalculated node-degree
 scenario}. Most out of range are the points at $c_{\rm s}=0.04$, $\gamma=2.72$ (Sao Paolo)
and at $c_{\rm s}=0.115$, $\gamma=(5.16)$ (Taipei).
 \label{fig7} }
\end{figure*}

Another useful observation concerns the correlation between the PTN
attack resilience and the node-degree distribution exponent $\gamma$
(\ref{2}). As we have observed in the previous studies
\cite{Ferber,Ferber09a} some of the PTNs under consideration are
scale-free: their node-degree distributions have been fitted to a
power-law decay (\ref{2}) with the exponents shown in Table
\ref{tab1}. Others are characterized rather by an exponential decay,
but up to a certain accuracy they can also be approximated by a
power-law behavior (then, the corresponding exponent is shown in
Table \ref{tab1} in brackets). In Fig. \ref{fig7}{\bf a} we show the
correlation between the fitted node-degree distribution exponent
$\gamma$ and $c_{\rm s}$ for the random attack scenario. Filled
circles correspond to scale-free PTNs, open circles correspond to
the PTNs where the scale-free behavior is less pronounced. It is
interesting to observe, that even if we include the PTNs which are
better described by the exponential decay of the node-degree
distributions, there is a notable tendency to find PTNs with smaller
values of $\gamma$ to be more resilient as indicated by larger
values of $c_{\rm s}$. This tendency is again confirmed if one
considers the recalculated node degree attack scenario, as shown in
Fig. \ref{fig7}{\bf b}.

\begin{figure*}
\centering
\includegraphics[width=80mm]{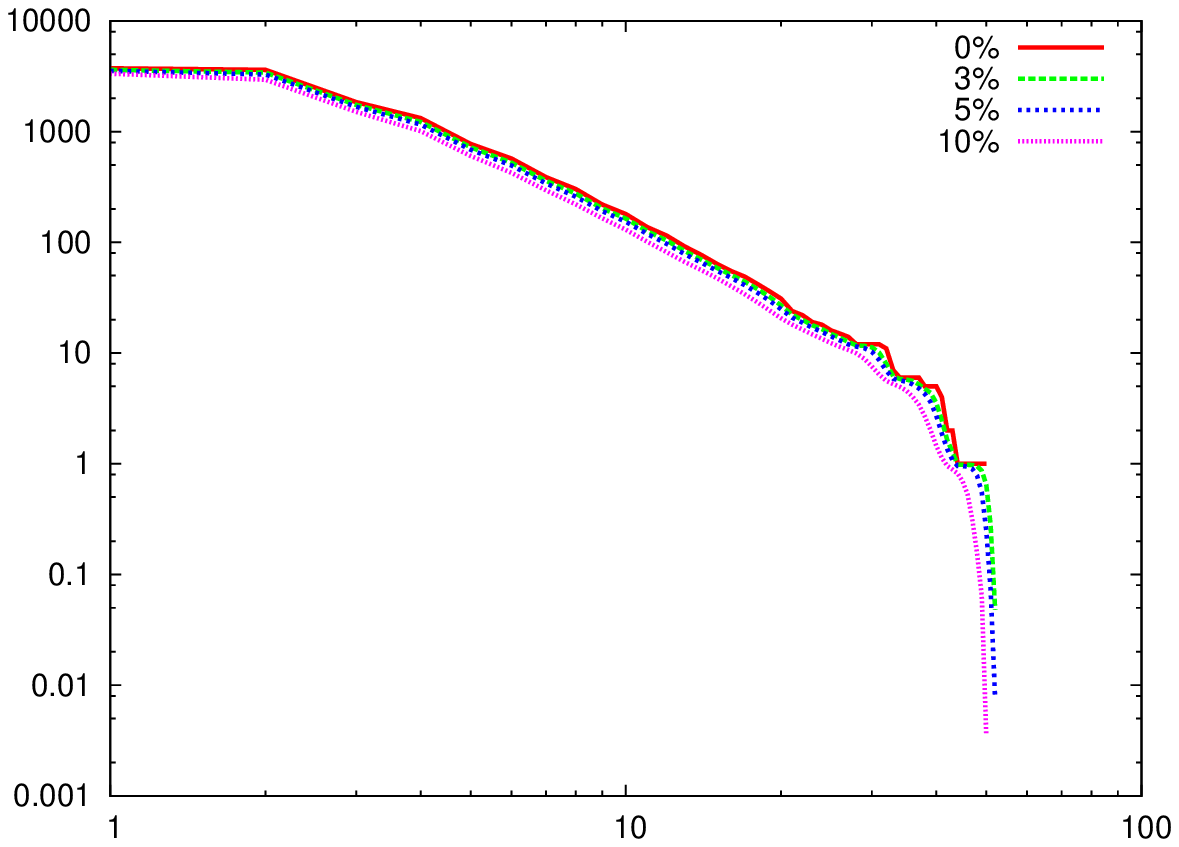} \hspace*{3em}
\includegraphics[width=80mm]{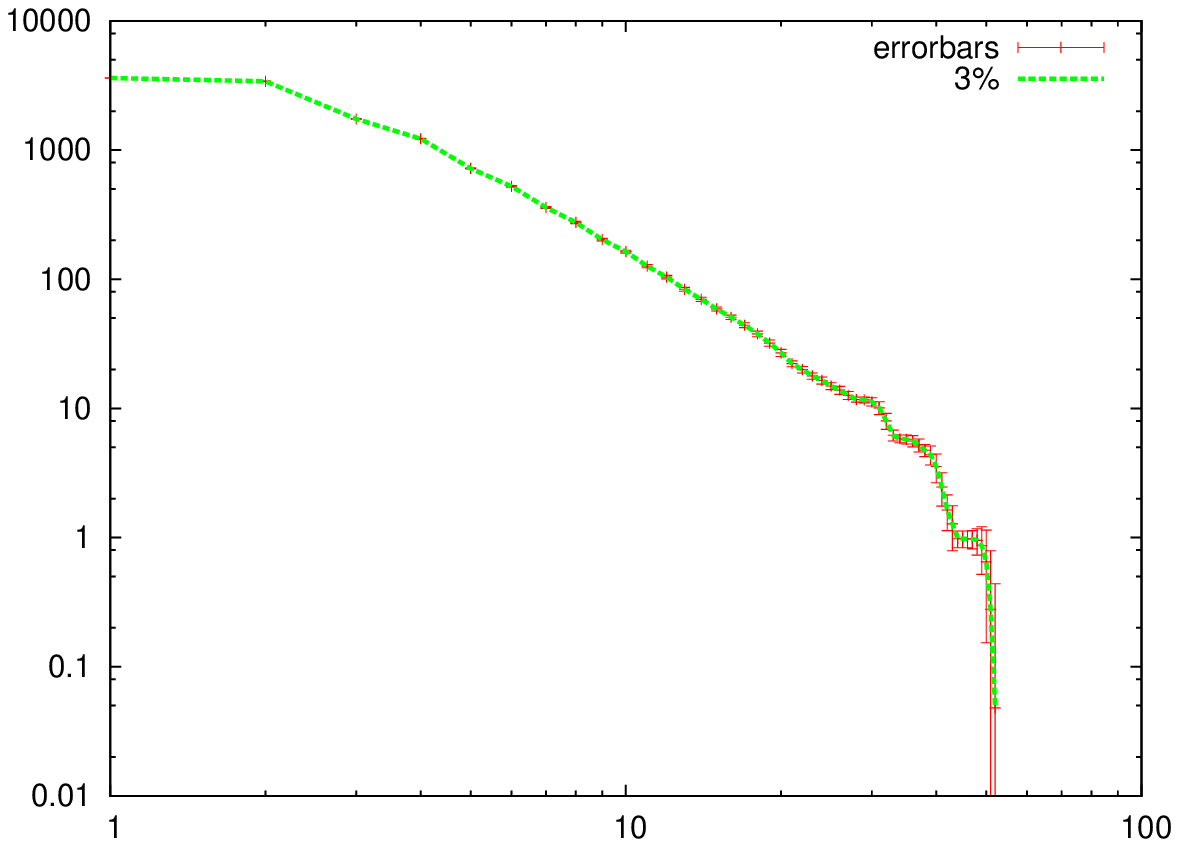}
\vspace*{3ex}
 \centerline{{\bf a.} \hspace*{30em} {\bf b.}}
 \vspace*{3ex}
\includegraphics[width=80mm]{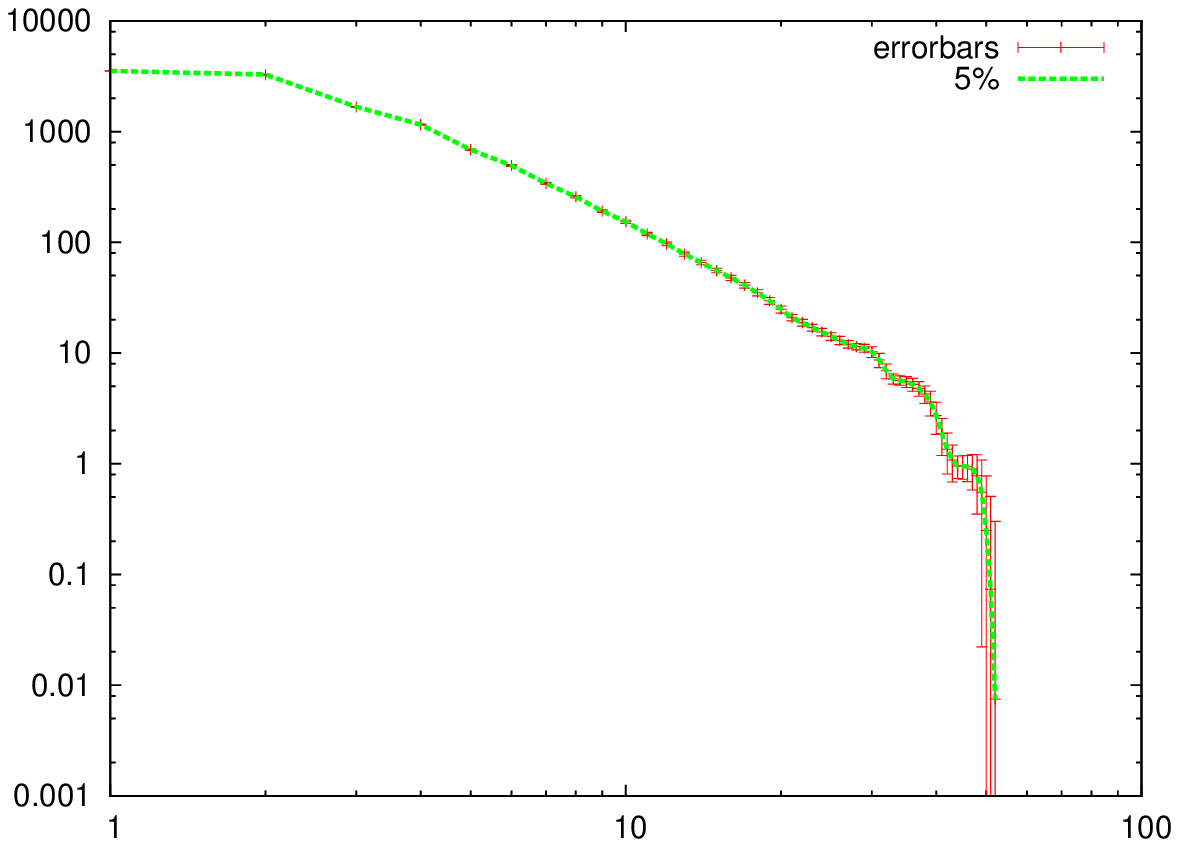} \hspace*{3em}
\includegraphics[width=80mm]{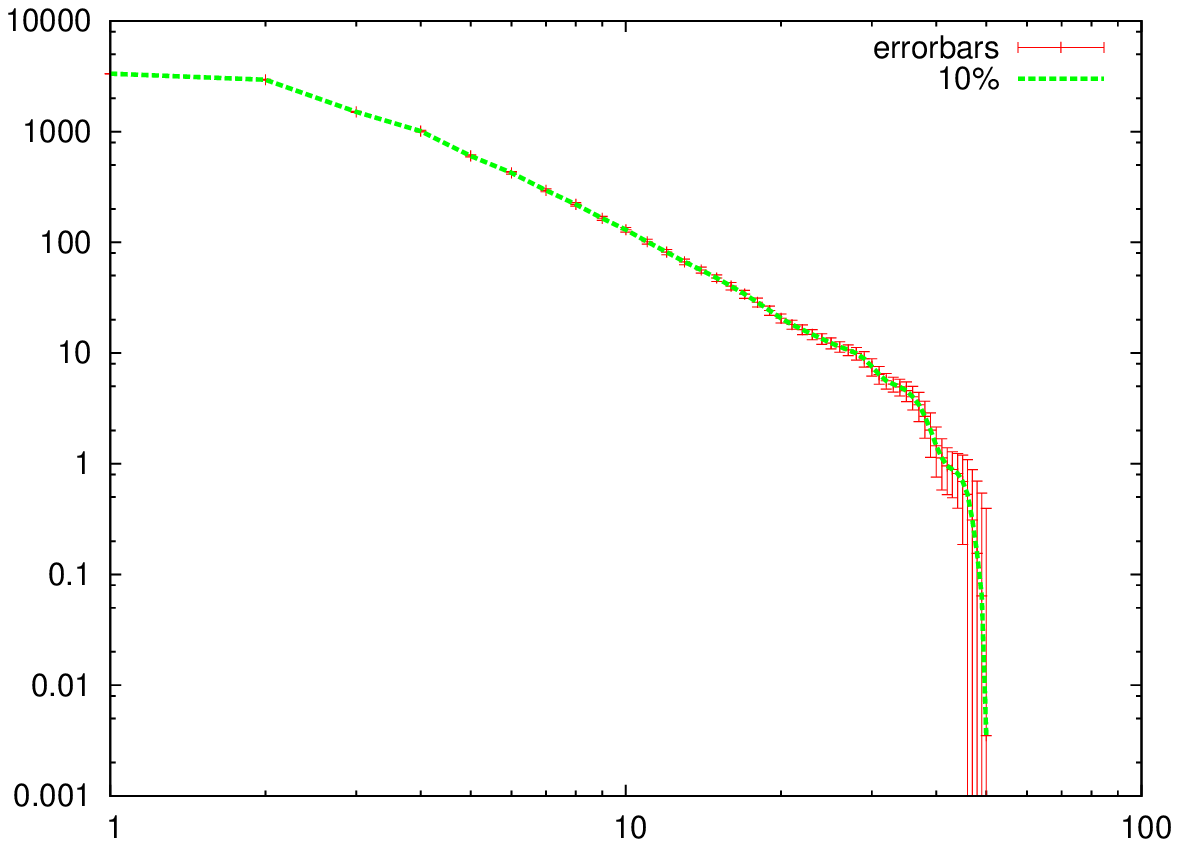}
\vspace*{3ex}
 \centerline{{\bf c.} \hspace*{30em} {\bf d.}}
 \caption{(color online). $\mathds{L}$-space.  Average cumulative node degree distributions for
 Paris PTN for the random attack scenario. Comparison
 of   the initial distribution (red curve, $c=0$) with those of the PTNs with
 $c=0.03$, $c=0.05$, $c=0.1$ ({\bf a}). Average cumulative node degree distribution together with
 statistical errors for $c=0.03$ ({\bf b}), $c=0.05$ ({\bf c}), $c=0.1$ ({\bf d}).}
 \label{fig8}
\end{figure*}

The above observed correlation between the exponent $\gamma$ that
characterizes the unperturbed network (i.e. a PTN at $c=0$) and the
segmentation concentration $c_{\rm s}$ at which however the PTN is
to a large part unperturbed indicates that some global properties of
the node-degree distribution may remain essentially unchanged when
the nodes are removed (i.e. a scale-free distribution remains
scale-free as $c$ increases, $0<c<c_{\rm s}$). To check that
assumption for the RV scenario, we analyzed the averaged cumulative
node degree distributions for each of the PTNs with 3,5, and 10 \%
of removed nodes. The cumulative distribution $P(k)$ is defined in
terms of the node-degree distribution $p(q)$ (\ref{2}) as:
\begin{equation}\label{15}
P(k)=\sum_{q=k}^{k^{\rm max}} p(q),
\end{equation}
with $k^{\rm max}$ the maximal node degree in the given PTN. Typical
results of this analysis are shown in Fig. \ref{fig8}, for the PTN
of Paris. We compare the cumulative node degree distribution $P(k)$
of the unperturbed PTN with that of the PTN where a given fraction
$c$ part of the nodes ($c=0.03$, 0.05, and 0.1, correspondingly) was
removed according to the random attack scenario (RV). For each of
the concentrations of the removed nodes, $P(k)$ was averaged over
2000 repeated attacks.

In the first plot, Fig. \ref{fig8}{\bf a}, we compare the three
resulting average distributions (for $c=0.03$, 0.05, and 0.1) with
the original one ($c=0$). One clearly sees that there is no
qualitative or even quantitative (change of exponent) change of the
distributions for any of the three cases. Indeed, if one has a large
set of nodes with a given node-degree distribution any sufficiently
large random subset of these nodes should have the same
distribution; in particular this holds if one averages these subset
distributions over many instances. The above argument seems to
ignore the change of degrees in the subset due to cutting off those
vertices not remaining in the set. However, due to the random choice
of the removed nodes the share of lost degree will on the average be
proportional to the degree of each vertex: the higher its degree the
more probable it is that one of its neighbors is chosen to be
removed and this probability is proportional to its degree. Thus,
the sum of degrees in the remaining subset is lower; but the degree
distribution $P(k)$ is effectively transformed to $P'(ck) = n P(k)$
where $c$ is the probability of any node being removed and $P'(k)$
is the distribution in the remaining subset of nodes, $n$ a
normalization. For an exponential distribution this transformation
shifts the scale. However, a scale free distribution keeps its
exponent under such a transformation.

In the other three plots, Figs. \ref{fig8}{\bf b}-{\bf d} we show
for each amount of removed nodes the average cumulative distribution
together with statistical errors calculated as the standard
deviation within the ensemble of the 2000 instances generated in the
sample. Even on the logarithmic scale these are very small for all
but the very high degrees where fluctuations of small numbers of
often less than one node for a given degree occur.

\section{Results in $\mathds{P}$-space}\label{IV}

Let us complement the $\mathds{L}$-space analysis performed above by
observing the reaction of PTN graphs under attack when one observes
them in another representation. In particular, we will investigate
$\mathds{P}$-space graphs.

\begin{figure*}
\centerline{\includegraphics[width=80mm]{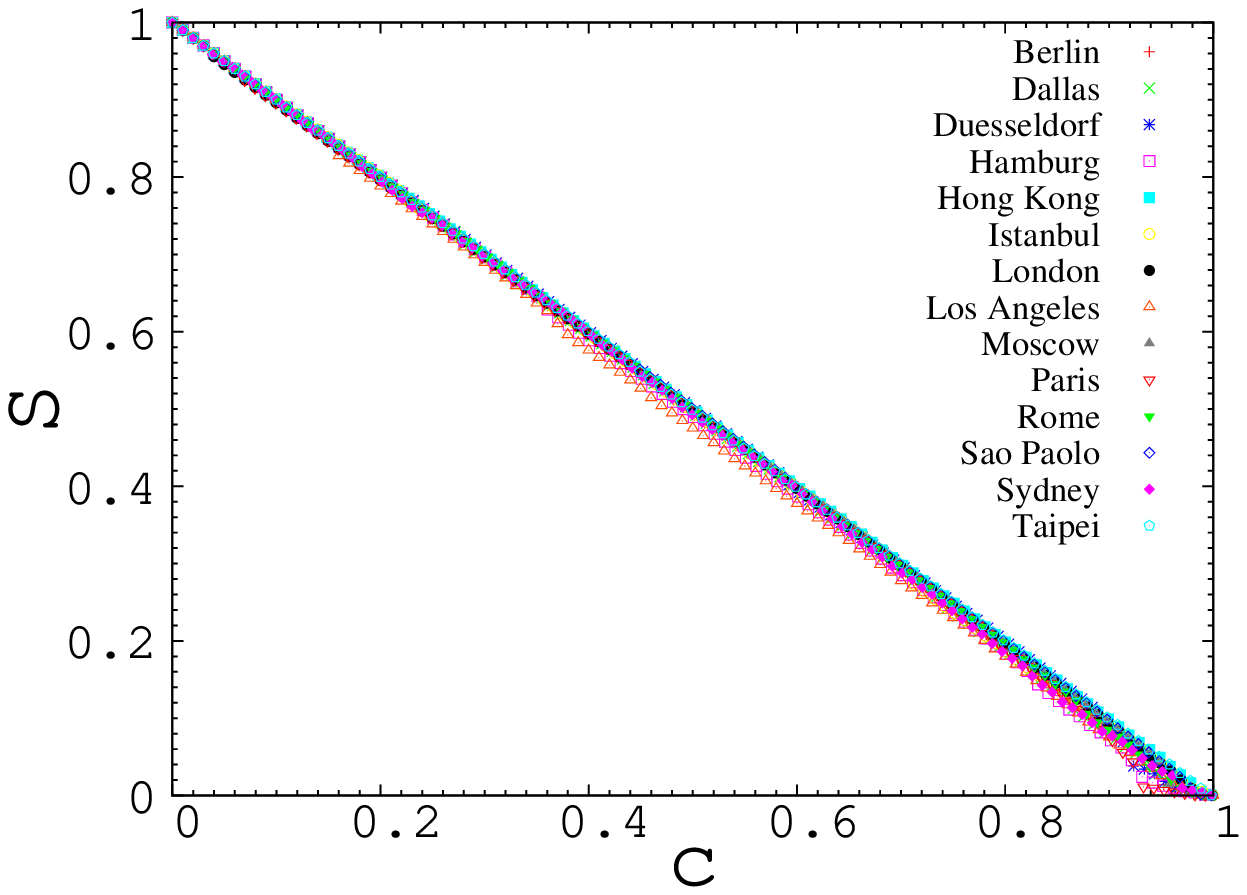}
\hspace*{3em}
\includegraphics[width=80mm]{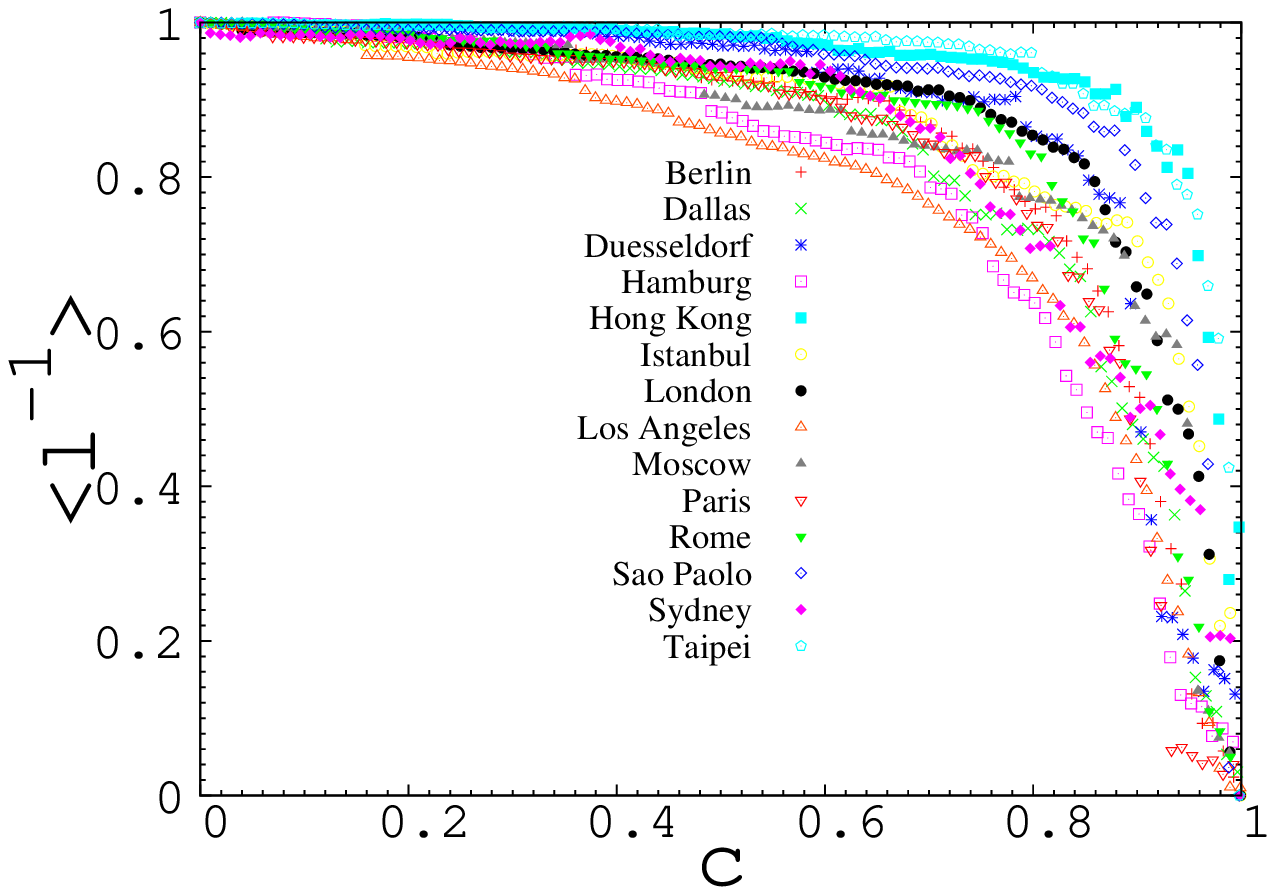}}
\centerline{{\bf a.} \hspace*{30em} {\bf b}}
 \caption{(color online). $\mathds{P}$-space. Random scenario. ({\bf a.}) size of the largest cluster $S$
 and ({\bf b.}) the average inverse
mean shortest path length $\langle\ell^{-1}\rangle$
 as functions of the fraction of
removed nodes $c$ normalized by their values at $c=0$. \label{fig9}}
\end{figure*}

First let us recall that in this representation each node
corresponds to a PTN station, i.e. it has the same interpretation as
in the $\mathds{L}$-space. However, the interpretation of a link
differs from that in the $\mathds{L}$-space: now all station-nodes
that belong to the same route are connected and thus each route
enters the $\mathds{P}$-space network as a complete subgraph. This
results in the main peculiarity of the interpretation of the
behavior under attacks of these graphs. Consider as an example the
$\mathds{P}$-space graph of Fig. \ref{fig1}{\bf c} and compare it to
the original PTN map, Fig. \ref{fig1}{\bf a}. Whereas the removal of
station node C in the map (Fig. \ref{fig1}{\bf a}) disconnects the
nodes B and D, the removal of the same node in the
$\mathds{P}$-space (Fig. \ref{fig1}{\bf c}) keeps nodes B and D
connected, as far as they still belong to the same route. Therefore,
the removal of nodes in $\mathds{P}$-space, performed either in a
random way or according to certain lists, has a different
interpretation in comparison to that occurring in the
$\mathds{L}$-space. An interpretation of the removal of nodes in
$\mathds{P}$-space is the following: if a node is removed, the
corresponding stop of the route is canceled while the route
otherwise keeps operating. If in the above example the station-node
C is removed, route No 2 still keeps operating and station-node B
can be reached from  D, only without stopping at C (e.g. the bus
takes a shortcut). In this way, as we will see below, the removal of
nodes in $\mathds{P}$-space allows us to gain additional insight
into the PTN structure.

\begin{figure*}
\centerline{\includegraphics[width=80mm]{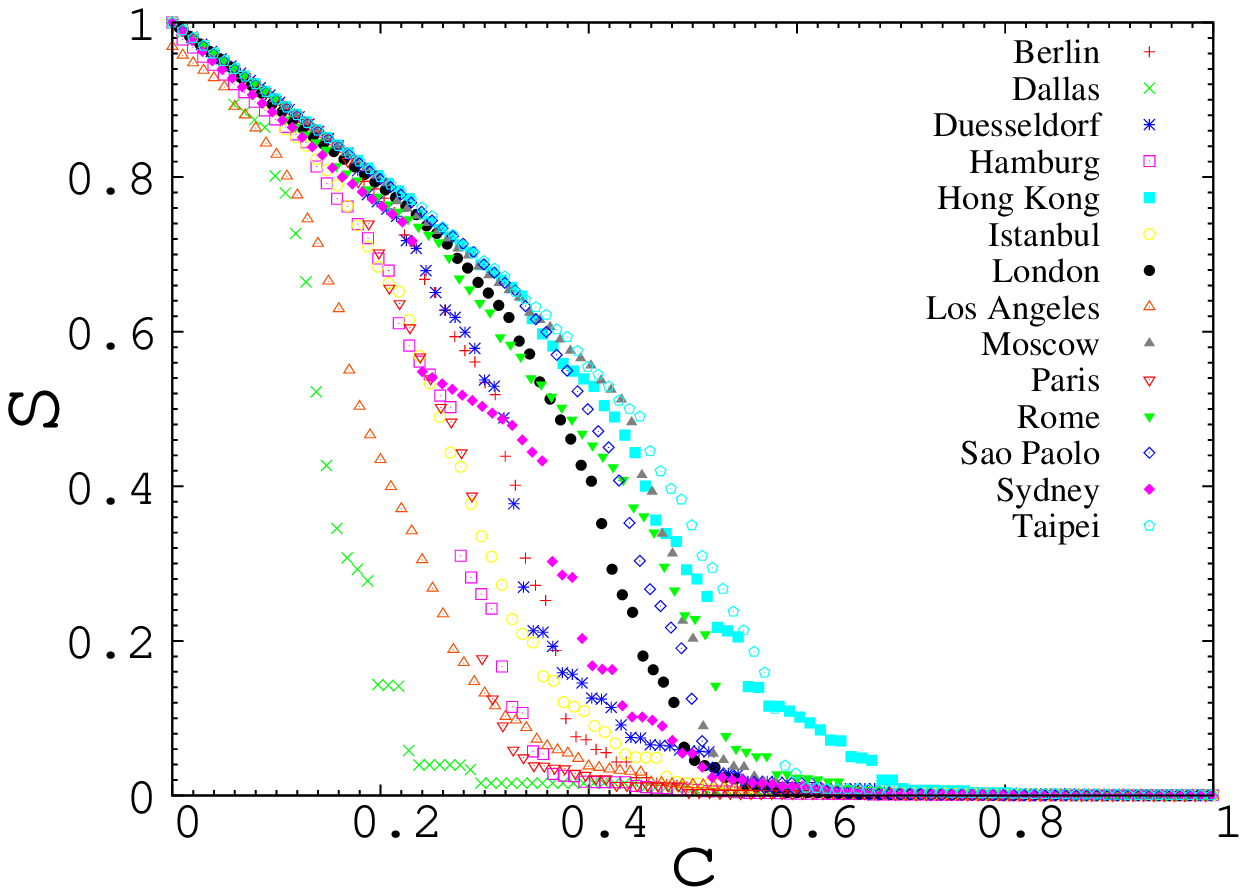}
\hspace*{3em}
\includegraphics[width=80mm]{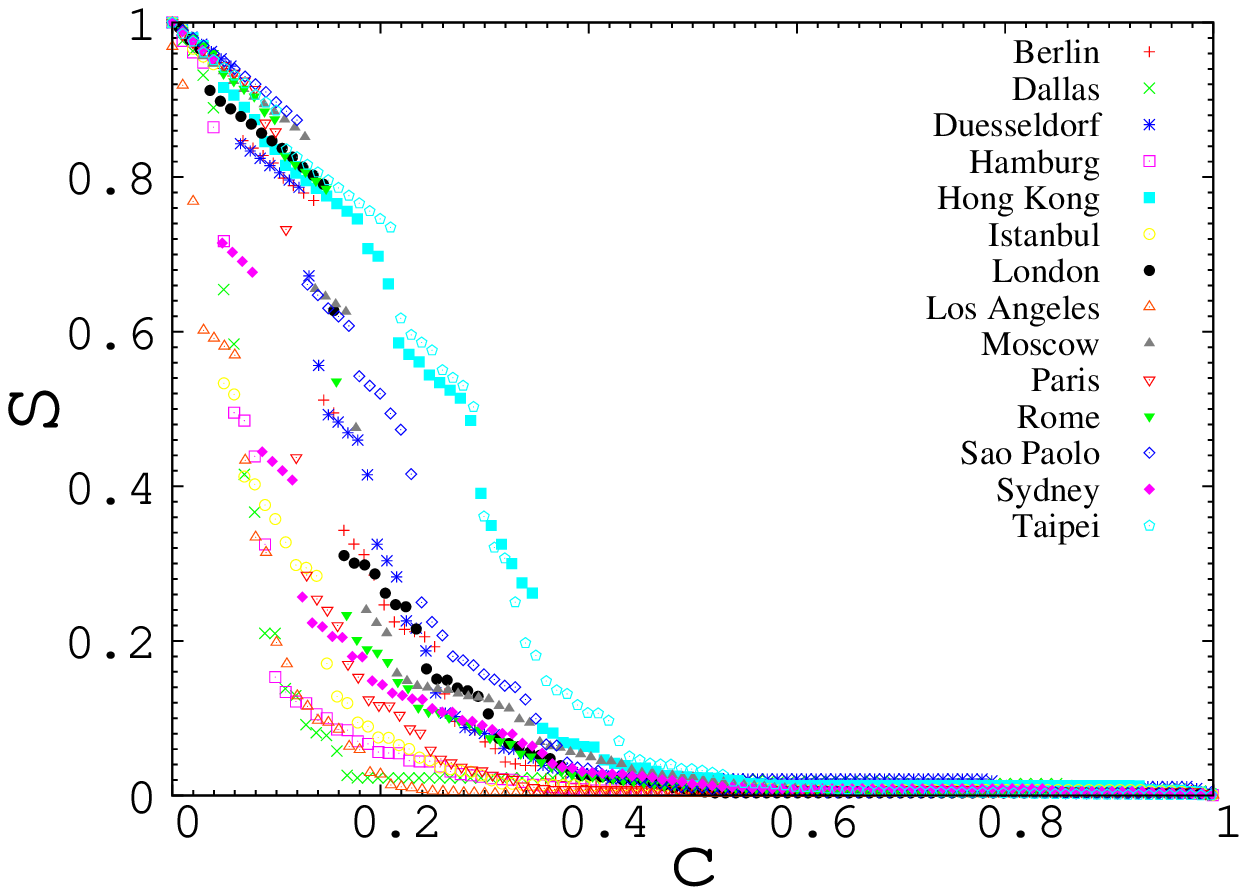}}
\centerline{{\bf a.} \hspace*{30em} {\bf b}}
 \caption{(color online). $\mathds{P}$-space, size of the largest cluster $S$ at   {\bf a}: highest degree scenario (recalculated), {\bf b}:
highest betweenness scenario (recalculated). \label{fig10}}
\end{figure*}

\begin{table*}[htbp]
\centering \tabcolsep2.5mm
\begin{tabular}{lllllllllllll}
\hline \hline
           City& $c_{\rm s}$ &  & $c_{\rm s}$ &  & $c_{\rm s}$ &  & $c_{\rm s}$ &     &  $c_{\rm s}$  &  & $c_{\rm s}$
\\ \hline
 Berlin        &   .155 & $C_B$  & .175  & $C_C$    & .215& $C_S$  & .285 & $C^{i}$& .290   &  $C^{i}_B$ &  .490                  & RV   \\
 Dallas        &   .065 & $C_B$  & .075    & $C_C$  & .095& $C_S$  & .115 & $C$    & .130        &  $C^{i}$     &  .490                        & RV   \\
 D\"usseldorf  &   .160 & $C_B$  & .185   & $C_S$   & .255& $C_C$  & .295 & $C^{i}$&  .300  &  $k^{i}$          &  .495         & RV   \\
 Hamburg       &   .050 & $C_C$  & .065    & $C_B$  & .145& $C_G$  & .170 & $C$    & .175       &  $C^{i}_C$             &  .490                 & RV   \\
 Hong Kong     &   .285 & $C_B$  & .295   & $C_S$   & .335& $C_C$  & .365 & $C$    & .380         &  $C^{i}$         &  .505               & RV   \\
 Istanbul      &   .060 & $C_C$  & .060   & $C_B$   & .060& $C^{i}_B$ & .115& $C^{i}_C$ & .175 &  $C$            &  .500              & RV   \\
 London        &   .155 & $C_B$  & .205   & $C_C$   & .305& $C_G$  & .330 & $C$    & .350         &  $C^{i}$          &  .495                 & RV   \\
 Los Angeles   &   .065 & $C_B$  & .095    & $C_C$  & .145& $C_S$  & .145 & $C^{i}_B$&  .150&  $C$       &  .480                   & RV   \\
 Moscow        &   .175 & $C_B$  & .255  & $C_C$    & .285& $C_S$  & .345 & $C$     & .395       &  $C^{i}$,$C^{i}_S$            &  .495& RV   \\
 Paris         &   .115 & $C_B$  & .165  & $C_S$    & .215& $C_C$  & .235 & $C^{i}_B$& .240 &  $C$,$C^{i}$    &  .500            & RV   \\
 Rome          &   .135 & $C_C$  & .160  & $C_B$    & .225& $C_G$  & .285 & $C_S$   &  .305    &  $C$           &  .495              & RV   \\
 Sa\~o Paolo   &   .205 & $C_B$,$C_C$ &.240& $C_S$  & .355& $C_G$  & .365 & $C$     & .390        &  $C^{i}$       &  .500           & RV   \\
 Sydney        &   .075 & $C_C$  & .085   & $C_B$   & .105& $C_S$  & .225 & $C$     &  .240        &  $C^{i}$         &  .510                  & RV   \\
 Taipei        &   .290 & $C_B$  & .320    & $C_S$  & .370& $C_C$  & .430 &  $C_G$  & .440 & $k,C_S^{i}$ &        .495                                                                               & RV   \\
 \hline \hline
\end{tabular}
\caption{Segmentation concentration $c_{\rm s}$ for different attack
scenarios applied to different PTNs in $\mathds{P}$-space.  For each
city, the Table shows the five most effective attack scenarios
ordered by increasing values of $c_{\rm s}$. The scenario is
indicated after corresponding value of $c_{\rm s}$. The scenarios
are abbreviated by the name of the characteristics used to prepare
the lists of removed nodes (see Sec. \ref{II} for detailed
explanation). In the last column the value of $c_{\rm s}$ for the
random scenario (RV) is shown. \label{tab3} }
\end{table*}

As in the case of the $\mathds{L}$-space representation, we study
the resilience of the $\mathds{P}$-space PTN graphs to attacks
performed following the sixteen different scenarios defined in
Section \ref{III}. In Fig. \ref{fig9} we show the change of the size
of the largest cluster $S$ ({\bf a}) and the average inverse mean
shortest path length $\langle\ell^{-1}\rangle$ ({\bf b}) under
random attacks (RV). If one compares this behavior with that
observed for the RV scenario in $\mathds{L}$-space (see Fig.
\ref{fig2}) one sees, that all PTNs under consideration react in a
much more homogeneous way. In $\mathds{L}$-space random attacks lead
to changes of the largest connected component $S$ that range from an
abrupt breakdown (Dallas) to a slow smooth decrease (Paris). In
$\mathds{P}$-space one observes for the same scenario only a
decrease of $S$ which corresponds to the number of removed nodes. No
break-down of this cluster occurs in this scenario. The value of
$S(c_{\rm s})$ defined by the condition (\ref{11}) is given in the
last column of Table \ref{tab3}. It is worth to note, that the
behavior of the mean inverse shortest path length  $\langle
\ell^{-1} \rangle$ as function of the fraction $c$ of disabled nodes
is also qualitatively different between the two RV scenarios in
$\mathds{L}$- (Fig. \ref{fig2}{\bf b}) and $\mathds{P}$- (Fig.
\ref{fig9}{\bf b}) spaces. In $\mathds{L}$-space $\langle \ell^{-1}
\rangle$ decreases in general faster than linearly indicating an
increase of the path length between the nodes as well as
partitioning of the network. In $\mathds{P}$-space $\langle
\ell^{-1} \rangle$ remains for a large part unperturbed as the nodes
of the complete subgraph remain essentially connected and the
shortest path lengths remain almost unchanged until only a small
fraction of the network remains.

To further detail the situation, similar as in Section \ref{III}, we
summarize in Table \ref{tab3} the outcome of the five most harmful
attack scenarios and compare those with the random attack scenario.
As it follows from the Table and as is further supported by Fig.
\ref{fig10}, the betweenness-targeted scenarios appear to be the
most harmful. Following this observation let us investigate the role
of the highest betweenness nodes: above all these are the nodes (and
not the highest-$k$ hubs) that control the PTN behavior under
attack. The $\mathds{P}$-space degrees of these high-betweenness
nodes do not essentially differ from those of the hubs, therefore
they cannot be easily distinguished from the other nodes during
attacks according to highest-$k$ scenario. To support this
assumption, let us recall that in the $\mathds{P}$-space
representation each route enters the overall network as a complete
subgraph, with all nodes interconnected. Removing nodes from a
complete graph does not lead to any segmentation. The decrease of
the normalized size of this graph will be given by the exact formula
$S=1-c$ (which is - almost - reproduced by the RV scenario, c.f.
Fig. \ref{fig9}{\bf a}). Under such circumstances a special role is
played by those nodes that join different complete graphs (different
routes). The removal of such nodes will separate different complete
routes and as a result may lead to network segmentation. Naturally,
being between different complete subgraphs such nodes are
characterized by  high centrality indices, as observed above.
Moreover, as far as their direct neighbors belong to different
complete graphs, these neighbors are not connected between each
other resulting in a lower value of the clustering coefficient $C$.
From Table \ref{tab3} one sees that attacks based on choosing nodes
with low-$C$ values are very effective in $\mathds{P}$-space.

\begin{figure}
\includegraphics[width=80mm]{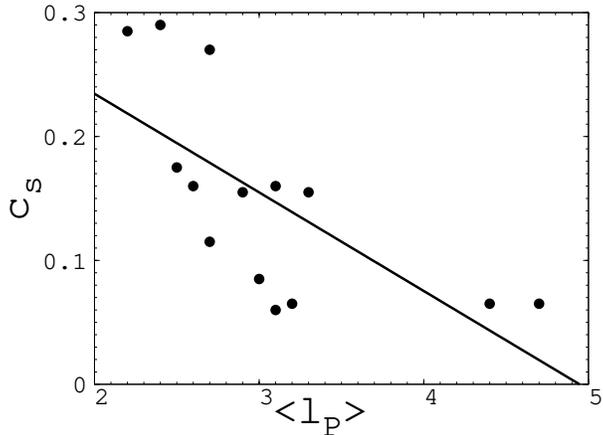}
 \caption{$\mathds{P}$-space. Correlations between the mean shortest path length $\langle \ell_{\mathds{P}} \rangle$
 and  segmentation concentration $c_{s}$ in the highest betweenness centrality scenario. The line serves as a guide to
observe the tendency of   $c_{s}$ to decrease with increasing
$\langle \ell_{\mathds{P}} \rangle$. \label{fig11}}
 \end{figure}

To conclude this section, we ask the question if a simple criterion
can be found that allows to predict a priori the $\mathds{P}$-space
PTN vulnerability. Namely, given the general PTN characteristics
(see Table \ref{tab1}) can one forecast resilience against attacks
in $\mathds{P}$-space? The answer is given by the observation that
the networks with low mean shortest path length $\langle
\ell_{\mathds{P}} \rangle$ are the best connected in
$\mathds{P}$-space and hence may be expected to be less vulnerable.
Indeed, on the one hand, for the above example of a complete graph
(a single PTN route) $\langle \ell_{\mathds{P}} \rangle=1$ and it is
extremely robust to $\mathds{P}$-space attacks. On the other hand, a
high value of $\langle \ell_{\mathds{P}} \rangle$ indicates numerous
intermediate nodes between different routes. As we have checked
above, the targeted removal of such nodes leads to rapid network
segmentation. In support of the above reasoning, in Fig. \ref{fig11}
we plot $c_{\rm s}$  as function of $\langle \ell_{\mathds{P}}
\rangle$ for attacks based on the highest betweenness centrality
scenario. There, within the expected scatter of data one observes a
clear evidence of the decrease of  $c_{s}$ with $\langle
\ell_{\mathds{P}} \rangle$, i.e. networks with higher mean path
length break down at smaller values of $c$ and are thus more
vulnerable.

It is worth to note here, that in $\mathds{P}$-space it is only the
RV attack  that has very similar impact on all PTNs (see Fig.
\ref{fig9}). As we have just observed, similar to the
$\mathds{L}$-space also in $\mathds{P}$-space the PTNs manifest
different level of robustness against attacks targeted on the most
important nodes. However, the order of vulnerability changes if one
compares the outcome of the $\mathds{L}$-space and
$\mathds{P}$-space attacks. This means that PTNs that were
vulnerable in the $\mathds{L}$-space may appear to be robust against
attacks in $\mathds{P}$-space. From Table \ref{tab3} we see that the
PTNs that are most stable against highest $C_B$-targeted attacks in
$\mathds{P}$-space are the PTNs of Hong Kong, Sa\~o Paolo, and
Moscow, with $c_{\rm s}= 0.285$, $0.205$, and $0.175$,
correspondingly. When attacked in $\mathds{L}$-space, the PTN of
Moscow keeps its robustness: $c_{\rm s}= 0.07$ during $C_B$-targeted
attack, which is one of highest $c_{\rm s}$ values for the
$\mathds{L}$-space, see Table \ref{tab2}. This is however not the
case for the PTNs of Hong Kong and Sa\~o Paolo. In
$\mathds{L}$-space, these belong to the most vulnerable PTNs.

\section{Conclusions and outlook}\label{V}

In this paper, we have studied the behavior of city public
transportation networks (PTNs) under attacks. In our analysis we
have examined PTNs of fourteen major cities of the world. The
principal motivation behind this study was to observe the behavior
under attack of a sample  of networks that were constructed for the
same purpose, to compare these with available analytical results for
percolation of complex networks, and possibly to derive some
conclusions about correlations between PTN characteristics
calculated a priory and the resilience to attacks. Furthermore, the
resilience behavior of a network against different attack scenarios
gives additional insight into the network architecture, discovering
structures on different scales. This approach has been termed the
'tomography' of a network \cite{Xulvi-Brunet03}.

In our study we have also attempted to compare our results with the
predictions of percolation theory on networks. Due to the sizes of
these systems which are far from the thermodynamic limit and the
rather small sample of networks no quantitative comparison appeared
possible. However, qualitative predictions about the location of
segmentation thresholds and thus the vulnerability could be
verified. Although our study was not primarily motivated by
applications, some of the results and methods developed within this
study may be useful for planning and risk assessment of PTNs. Our
analysis has identified PTN structures  which are especially
vulnerable and others, which are particulary resilient against
attacks. Further investigation of other relevant network properties
may reveal mechanisms behind this structural resilience
\cite{Holovatch09}. Furthermore we note that  the methods developed
here also allow to identify minimal strategies to obstruct the
operation of the PTN of a city e.g. for the purposes of industrial
action and possibly achieve a successful end of a social conflict.

To analyze PTN resilience we have applied different attack
scenarios, that range from a random failure to a targeted
destruction, when the most influential network nodes were removed
according to their operating characteristics. To choose the most
influential nodes, we have used different graph theoretical
indicators and determined in such a way the most effective attack
scenarios. By our paper we show that even within a sample of
networks that were created for the same purpose one observes
essential diversity with respect to their behavior under attacks of
various scenarios. Results of our analysis show that PTNs
demonstrate  rich variety of behavior under attacks, that range from
smooth decay to abrupt change.

As shown by our study, the impact of attacks may be measured by
different quantities. As a criterion that is well defined and easily
reproducible we choose to define the segmentation concentration
$c_{\rm s}$ to correspond to the situation where the largest
remaining cluster contains one half of the original nodes of the
network. Let us note as well, that definitely not all of the PTNs
analyzed demonstrated scale-free behavior in $\mathds{P}$-space (and
even less in $\mathds{L}$-space). Nevertheless, in spite of the
diversity of behavior we clearly see common tendencies in their
reaction to attacks. In particular, this enabled us to propose
criteria that allow an a priori estimate of PTN robustness. In
$\mathds{L}$-space resilience is indicated by a high value of the
Molloy-Reed parameter $\kappa$, Eqs. (\ref{1}), (\ref{12}) or by a
small value of the exponent $\gamma$, if a power law is observed for
the PTN node degree distribution, in $\mathds{P}$-space high
resilience is indicated by a small mean shortest path length
$\langle \ell_{\mathds{P}} \rangle$.

One of possible continuations of our study will be the analysis of
PTN resilience in other graph representations, than those that were
described above.

\section*{Acknowledgements}

CvF wishes to thank Reinhard Folk for his hospitality at the
Institute of Theoretical Physics, JKU Linz. This work was supported
by the Austrian Fonds zur F\"orderung der wissenschaftlichen
Forschung under Project No. P19583-N20 and by the cooperation
programme 'Dnipro' between the  Ministry of Foreign Affairs of
France  and the Ministry of Education and Science of Ukraine.


\begin{thebibliography}{50}
\bibitem{note2} Note however that an early paper on metabolism networks \cite{Jeong01} checked
43 different organisms.
\bibitem{note3} In the following we will only consider the removal of
nodes. This is along the lines of the site percolation phenomenon.
\bibitem{note4} Another obvious dfference between the  relative size
of the largest cluster $S$  of a network considered here and a
spanning cluster size on a regular lattice is that the former can be
zero only in the infinite network limit, whereas the latter can be
zero even for a finite size system.
\end{thebibliography}
\end{document}